\documentclass[12pt,english,floatfix,superscriptaddress,aps,prd,preprint,showkeys]{revtex4}
\usepackage{amsmath}
\usepackage{amssymb}
\usepackage{amsbsy}
\usepackage{amsfonts}
\usepackage{amsopn}
\usepackage{amstext}
\usepackage{graphicx}
\usepackage{amssymb}
\usepackage{amsfonts}
\usepackage{amsmath}
\usepackage{graphicx}
\usepackage[english]{babel}
\usepackage{color}
\usepackage{slashed}
\usepackage{esint}
\usepackage[dvips]{epsfig}
\usepackage[dvips]{graphicx}
\usepackage{float}
\usepackage{units}
\usepackage{textcomp}

\usepackage{hyperref}
\usepackage{slashed}
\usepackage{hyperref}
\usepackage{graphicx}
\usepackage{amssymb}
\usepackage{amsmath}
\usepackage{color}

\begin{document}


\title{Thick brane in f(T,B) gravity}


\author{A.R.P. Moreira}
\email{allan.moreira@fisica.ufc.br}
\affiliation{Universidade Federal do Cear\'a (UFC), Departamento de F\'isica,\\ Campus do Pici, Fortaleza - CE, C.P. 6030, 60455-760 - Brazil.}

\author{J.E.G. Silva}
\email{euclides.silva@ufca.edu.br}
\affiliation{Universidade Federal do Cariri(UFCA), Av. Tenente Raimundo Rocha, \\ Cidade Universit\'{a}ria, Juazeiro do Norte, Cear\'{a}, CEP 63048-080, Brazil}

\author{F.C.E. Lima} 
\email{cleiton.estevao@fisica.ufc.br}
\affiliation{Universidade Federal do Cear\'a (UFC), Departamento de F\'isica,\\ Campus do Pici, Fortaleza - CE, C.P. 6030, 60455-760 - Brazil.}


\author{C.A.S. Almeida}
\email{carlos@fisica.ufc.br}
\affiliation{Universidade Federal do Cear\'a (UFC), Departamento de F\'isica,\\ Campus do Pici, Fortaleza - CE, C.P. 6030, 60455-760 - Brazil.}

\begin{abstract}
In this paper we explore the five-dimensional $f(T,B)$ 
teleparallel modified gravity with $f_1(T,B)=T+k_1B^{n_1}$ and $f_2(T,B)=B+k_2T^{n_2}$ in the brane scenario. Asymptotically, the bulk geometry converges to an $AdS_5$ spacetime whose cosmological constant is produced by parameters that control torsion and boundary term.  The analysis of the  energy density condition reveals a splitting brane process satisfying the weak and strong energy conditions for some values of the parameters $n_{1,2}$ and $k_{1,2}$. In addition, we investigate the behavior of the gravitational perturbations in this scenario. In the bulk, the torsion keeps a gapless non-localizable and stable tower of massive modes. Inside the brane core the torsion produces new barriers and potential wells leading to small amplitude massive modes and a massless mode localized for some values of the parameters $n_{1,2}$ and $k_{1,2}$.
\end{abstract}

\keywords{Braneworld model, Modified theories of gravity, Boundary term, Teleparallelism.}

\maketitle

\section{Introduction}

Randall-Sundrum (RS) brane models \cite{rs,rs2}, have attracted much attention since they were proposed, because of their great success in solving the hierarchy problem \cite{rs2} and to enables a new approaches to address of  outstanding issues such as thethe cosmological problem \cite{cosmologicalconstant}, the nature of dark matter \cite{darkmatter} and dark energy. However, the thickness of the branes in the RS model is disappearing. A new approach is the so-called thick brane model, where the brane is constructed by
the scalar field \cite{Goldberger1999,Gremm1999, DeWolfe1999,Bazeia2008,Dzhunushaliev2009}. Various thick brane models have been investigated in Refs. \cite{Charmousis2001, Arias2002, Barcelo2003, Bazeia2004, CastilloFelisola2004, Navarro2004, BarbosaCendejas2005, Bazeia2007, Koerber2008,deSouzaDutra2008,conifold,Almeida2009, Cruz2013, Liu2011, Dutra2014}. All of these works only considered the contribution of spacetime curvature without torsion. However, torsion can also lead to the equivalent gravity theory known as the teleparallel equivalent of general relativity (TEGR) \cite{Hayashi1979, deAndrade1997, deAndrade1999, Aldrovandi}.

The TEGR  is constructed using the curvature-less tensor named Weitzenböck  instead  of the torsion-less Levi-Civita connection.  Furthermore, the fundamental dynamical quantity of the theory is not the metric tensor but the so-called, tetrad field. In such a formulation the gravitational Lagrangian results from contractions of the torsion tensor and is called the torsion scalar $T$, similarly to the General Relativity (GR) Lagrangian, i.e. the curvature scalar $R$, which is constructed by contractions of the curvature tensor. The torsion scalar and the curvature scalar are related by a boundary quantity, $B$, where $R=-T+B$.

Hence, similarly to the $f(R)$ extensions of GR \cite{DeFelice2010, Nojiri2011}, one can construct $f(T)$ extensions of TEGR \cite{Ferraro2007, Ferraro2011}. The interesting feature in this extension is that $f(T)$ does not coincide with $f(R)$ gravity, despite the fact that TEGR coincides with GR. Additionally, the advantage of this theory is that the equations of motion are of second order, in contrast to the fourth-order equations of $f(R)$ gravity.  Since it is a new gravitational modification class, various thick brane models have been investigated in Refs. \cite{Yang2012, Capozziello, Menezes, tensorperturbations, ftnoncanonicalscalar, ftborninfeld, ftmimetic}.

New teleparallel gravity models have emerged, such as the $f(T, T_G)$ gravity, where $T_G$  is the torsion scalar equivalent of Gauss-Bonnet (GB) \cite{Kofinas2014,Kofinas2014a,Chattopadhyay2014}, the $f(T,\mathcal{T})$ gravity, where $\mathcal{T}$ is the trace of the stress-energy tensor \cite{Saez-Gomez2016}, and $f(T,B)$ gravity, where $B$ is the boundary term \cite{Bahamonde2015, Wright2016, Bahamonde2016}. Among these  models, the gravitational model $f(T,B)$ it is more interesting, due this model features have less mathematical complexity as well as good agreement with observational data to describe the accelerated expansion of the universe \cite{Franco2020,EscamillaRivera2019}. In addition, significant results were obtained in cosmological perturbations and thermodynamics, and dark energy, and gravitational waves \cite{Bahamonde2016a, Caruana2020,Pourbagher2020,Bahamonde2020a,Azhar2020,Bhattacharjee2020,Abedi2017}.   This model  provides equivalence between torsion and curvature, where one simultaneously recovers both the models of $f(T)$ gravity and $f(R)=f(-T+B)$ gravity. 

Considering the increasing interest in modified teleparallel gravity models and in the significant results obtained in $f(T,B)$ gravity that gives us the possibility to treat models alternatives to GR, in this paper we investigate the impact of torsion and boundary term on the structure of branes and their influences in the localization of gravity on the  branes.

Assuming only one scalar field as source we obtained a splitting in the thick  brane. By varying the parameters that control the  torsion and boundary term, the source undergoes a phase transition revealed by the energy density components. The gravitational perturbations form a gapless Kaluza-Klein (KK) spectrum whose interaction with the brane depends on the torsion and boundary term parameters.

The paper is organized as follows. In section (\ref{sec1}) we review the main definitions of the teleparallel  theory, we introduce the $f(T,B)$ theory and then give the field equations for the five-dimensional braneworld. Furthermore, the exterior and interior solutions are found. We examine the energy density behavior in the brane. In section (\ref{sec2}) we derive the tensor perturbed equations and explore the gravitational KK modes. Finally, additional comments are presented in section (\ref{sec3}).

\section{Modified Teleparallel braneworld}
\label{sec1}

In this section we present the main concepts of the modifield teleparallel $f(T,B)$ gravity and obtain the modified gravitational equations for the braneworld scenario.

In teleparallel gravity, the dynamic variable is provided by the vielbeins, defined by $g_{MN}=\eta_{ab}h^a\ _M h^b\ _N$, where the capital latin index $M={0,...,D-1}$ denotes the bulk coordinate indexes and the latin index $a={0,...,D-1}$ is a vielbein index.

In order to allow a distant parallelism, the teleparallel gravity assumes a curvature free connection, known as the Weitzenb\"{o}ck connection, defined by  
\cite{Aldrovandi}
\begin{eqnarray}\label{a.685}
\widetilde{\Gamma}^P\ _{NM}=h_a\ ^P\partial_M h^a\ _N.
\end{eqnarray}
The Weitzenb\"{o}ck connection has a non vanishing torsion, defined as
\begin{eqnarray}
T^{P}\  _{MN}= \widetilde{\Gamma}^P\ _{NM}-\widetilde{\Gamma}^P\ _{MN}.
\end{eqnarray}
The Weitzenb\"{o}ck and the Levi-Civita connections are related by $\widetilde{\Gamma}^P\ _{NM}= \Gamma^P\ _{NM} + K^P\ _{NM}$
where 
\begin{eqnarray}
K^P\ _{NM}=\frac{1}{2}\Big( T_N\ ^P\ _M +T_M\ ^P\ _N - T^P\ _{NM}\Big)
\end{eqnarray}
 is the contorsion tensor \cite{Aldrovandi}.

By defining the so-called superpotential torsion tensor as 
\begin{eqnarray}
S_{P}\ ^{MN}=\frac{1}{2}\Big( K^{MN}\ _{P}-\delta^N_P T^{QM}\ _Q+\delta^M_P T^{QN}\ _Q\Big),
\end{eqnarray} 
the torsion scalar is
\begin{eqnarray}
T=\frac{1}{2} T^{P}\  _{MN} T_{P}\ ^{MN} +T^{P}\ _{MN}T^{NM}\ _{P}-2T^{P}\ _{MP}T^{NM}\ _{N}=T_{PMN}S^{PMN}.
\end{eqnarray} 
Therefore, the Lagrangian of TEGR reads
\begin{eqnarray}
\mathcal{L}=- hT/4\kappa_g ,
\end{eqnarray}
where $h=\sqrt{g}$, with $g$ the determinant of the metric and $\kappa_g=4\pi G/c^4$ is the gravitational constant \cite{Aldrovandi}. Such Lagrangian is equivalent to the usual Einstein-Hilbert action. Indeed, the Ricci scalar for the Weitzenb\"{o}ck is proportional to $T$ by \cite{Aldrovandi}
\begin{eqnarray}
R=-T-2\nabla^{M}T^{N}\ _{MN},
\end{eqnarray}
and thus we can identify the boundary term 
\begin{eqnarray}
B\equiv -2\nabla^{M}T^{N}\ _{MN}=\frac{2}{h}\partial_M(h T^M),
\end{eqnarray}
in which $T^M$ as the torsion tensor that can
be define by $T_M=T^N\ _{MN}$. Hence, one can immediately see that GR and TEGR will lead to exactly the same equations. However,
this will not be the case if one uses $f(R)$ or $f(T)$ as the
Lagrangian of the theory, which therefore corresponds to
different gravitational modifications \cite{Abedi2017}.

A modified gravity theory can be accomplished by considering as the gravitational Lagrangian a function of $T$ and $B$, leading to $f(T,B)$ gravity \cite{Abedi2017}. We assume a five-dimensional $f(T,B)$ gravity in the form  
\begin{eqnarray}\label{55.5}
\mathcal{S}=-\frac{1}{4\kappa_g}\int h f(T,B)d^5x+\int \left(\Lambda +\mathcal{L}_m\right)d^5x,
\end{eqnarray}
where $\mathcal{L}_m$ is the matter Lagrangian.
By varying the action with respect to the vierbein we
obtain the following field equations \cite{Abedi2017,Pourbagher2020}
\begin{eqnarray}\label{3.36}
\frac{1}{h}f_T\Big[\partial_Q(h S_N\ ^{MQ})-h\widetilde{\Gamma}^R\ _{SN}S_R\ ^{MS}\Big]+\frac{1}{4}\Big[f-Bf_B\Big]\delta_N^M& &\nonumber\\+\Big[(\partial_Qf_T)+(\partial_Qf_B) \Big]S_N\ ^{MQ} +\frac{1}{2}\Big[\nabla^M\nabla_N f_B-\delta^M_N\Box f_B\Big]&=&-\kappa_g(\Lambda\delta_N^M+\mathcal{T}_N\ ^M),
\end{eqnarray}
where $\Box\equiv\nabla^M\nabla_M$, $f\equiv f(T,B)$, $f_T\equiv\partial f(T,B)/\partial T$ e $f_{B}\equiv\partial f(T,B)/\partial B$  and $\mathcal{T}_N\ ^M$ is the stress-energy tensor, which in terms of
the matter Lagrangian is given by $\mathcal{T}_a\ ^M=-\delta\mathcal{L}_m/\delta h^a\ _M$.

We would like to consider the braneworld scenario,
for which the metric ansatz reads as \cite{Yang2012}

\begin{equation}\label{45.a}
ds^2=e^{2A(y)}\eta_{\mu\nu}dx^\mu dx^\nu+dy^2,
\end{equation}
where $\eta_{\mu\nu}=(-1, 1, 1, 1)$ is the four-dimensional
Minkowski metric and $e^{A(y)}$ is the so-called warp factor. Accordingly, adopting the \textit{sechsbeins} in the form $h^a\ _M=diag(e^A, e^A, e^A, e^A, 1)$, the torsion scalar and the boundary term are obtained respectively as
\begin{eqnarray}
 T&=&-12A'^2,\nonumber\\
 B&=&-8(A''+4A'^2),
\end{eqnarray}
which together reproduce the Ricci scalar $R=-T+B=-4(5A'^2+2A'')$, where the prime $(\ '\ )$ denotes differentiation with respect to $y$. 

In our model, we take a Lagrangian given by
\begin{equation}
\mathcal{L}_m=-h\left[\frac{1}{2}\partial^M\phi\partial_M\phi+V(\phi)\right],
\end{equation}
where $\phi\equiv \phi(y)$ is a background scalar
field that generates the brane. The explicit field equations (\ref{3.36}) and the equation of motion of
the scalar field can be written as

\begin{eqnarray}
\phi''+4A'\phi'&=&\frac{\partial V}{\partial \phi},\label{scalarfieldeom}\\
\frac{1}{4}\Big[f+8(A''+4A'^2)f_B\Big]+6A'^2f_T&=&-\kappa_g\Big(\Lambda-\frac{1}{2}\phi'^2+V \Big),\label{e.1}\\
-12\Big[A'(A'''+8A'A'')(f_{BB}+f_{TB})+3A''A'^2(f_{TT}+f_{BT})\Big]& &\nonumber\\
+\frac{1}{2}(A''+4A'^2)(4f_{B}+3f_{T})+\frac{1}{4}f&=& -\kappa_g\Big(\Lambda+\frac{1}{2}\phi'^2+V \Big).\label{e.2}
\end{eqnarray}

The equations (\ref{scalarfieldeom}), (\ref{e.1}) and (\ref{e.2})  form a quite intricate system of coupled equations. Thus, we first analyse the solutions exterior to the brane and then, we propose a possible solution for the brane core.

In this work we consider a power-law modified gravity in the form $f_1(T,B)=T+k_1B^{n_1}$ and $f_2(T,B)=B+k_2T^{n_2}$ , where $k_{1,2}$ and $n_{1,2}$ are parameters controlling the deviation of the usual teleparallel theory.
  
\subsection{Thin brane regime}
Let us first explore the effects of the torsion on the vacuum region, exterior to the  brane. In the vacuum, the stress-energy tensor vanishes $\mathcal{T}_N\ ^M=0$ and then, the geometry is governed by the bulk cosmological constant $\Lambda$.

Assuming a vacuum solution $A'=-c$ that establishes a relation between the bulk cosmological constant, the torsion parameters and $c$, the corresponding exterior metric takes the form 
\begin{eqnarray}\label{a.65}
ds^2=e^{-2c|y|}\eta_{\mu\nu}dx^\mu dx^\nu + dy^2,
\end{eqnarray}
which is the thin brane solution in the RS model \cite{rs,rs2}. However, unlike the usual GR braneworld, the $f(T,B)$ gravity enables new possible configurations. We give two examples below.

\subsubsection{$f_1(T,B)=T+k_1B^{n_1}$}
The modified gravitational field equations (\ref{e.1}) and (\ref{e.2}) yields to

\begin{eqnarray}\label{e.1.a}
-3c^2+(-4)^{n_1-1}k_1(n_1-1)(8c^2)^{n_1}=\kappa_g\Lambda.
\end{eqnarray}
Firstly, let us seek for solutions of Eq.(\ref{e.1.a}) for $\Lambda=0$. If $n_1=1$ the only solution is $c=0$, which leads to a factorizable Kaluza-Klein model. Yet, for $n_1=2$ we obtain the solution
\begin{eqnarray}
\label{cosmologicalconstant1}
c=\pm \sqrt{\frac{3\kappa_g}{256 k_1}},
\end{eqnarray}
whereas for $n=3$ we find 
\begin{eqnarray}
c=\pm \sqrt[4]{\frac{3\kappa_g}{8192k_1}}.
\end{eqnarray} 
Accordingly, the $f_1(T,B)$ yields to a warped compactified spacetime even in the absence of a bulk cosmological constant. 

Now let us consider a non-zero bulk cosmological constant.
For $n_1=1$, the Eq.(\ref{e.1.a}) yields to
\begin{eqnarray}
c= \pm\sqrt{\frac{\kappa_g\left(-\Lambda\right)}{3}}\quad,
\end{eqnarray}
which are solutions that represent the usual thin brane with an $AdS_5$ bulk.
For $n_1=2$, we obtain four solutions
\begin{eqnarray}
c= \pm\sqrt{\frac{\kappa_g\left(-\Lambda\right)}{3}}\quad or \quad  c= \pm\sqrt{\frac{\kappa_g\left(\Lambda+1\right)}{256k_1}}.
\end{eqnarray}
The last two solutions are generated by a positive cosmological constant provided $k>0$. 
For $n_1=3$ we obtain six solutions
\begin{eqnarray}
c= \pm\sqrt{\frac{\kappa_g\left(-\Lambda\right)}{3}}\quad or \quad  c= \pm\sqrt[4]{\frac{\kappa_g\left(-\Lambda-1\right)}{8192k_1}}.
\end{eqnarray}
The last solutions are generated by a positive cosmological constant provided $k<0$. 
Thus, for both $n_1=3$, $n_1=2$ and $n_1=1$, the exterior geometry is the same as in the thin RS model. 

\subsubsection{$f_2(T,B)=B+k_2T^{n_2}$}
The modified gravitational field equations (\ref{e.1}) and (\ref{e.2}) yields to
\begin{eqnarray}\label{e.1.b}
(-4)^{n_2-1}k_2(2n_2-1)(3c^2)^{n_2}=\kappa_g\Lambda.
\end{eqnarray}
For any $n_2$, the only solutions of Eq.(\ref{e.1.b}) with $\Lambda=0$ is $c=0$. Now let us consider a non-zero bulk cosmological constant. For $n_2=1$, the Eq.(\ref{e.1.b}) yields to
\begin{eqnarray}
c= \pm\sqrt{\frac{\kappa_g\left(-\Lambda\right)}{3k_2}}\quad.
\end{eqnarray}
If $k_2=1$, these solutions represent the usual thin brane with an $AdS_5$ bulk.
For $n_2=2$, we obtain solutions
\begin{eqnarray}
c= \pm\sqrt[4]{\frac{\kappa_g\left(\Lambda\right)}{108k_2}},
\end{eqnarray}
which are generated by a positive cosmological constant provided $k_2>0$. 
For $n_2=3$ we obtain six solutions
\begin{eqnarray}
 c= \pm\sqrt[6]{\frac{\kappa_g\left(-\Lambda\right)}{2160k_2}}.
\end{eqnarray}
The last solutions are generated by a positive cosmological constant provided $k_2<0$.  
The  solutions for any $n_2$ reveals a striking effect of  $f_2(T,B)$ upon the exterior geometry.

\subsection{Thick brane regime}

We can rewrite equations (\ref{e.1}) and (\ref{e.2}) as
\begin{eqnarray}
6A'^2&=&-\frac{\kappa_g}{f_T}\Big(\Lambda+P+P_{TB} \Big),\label{0.00005}\\
3A''+12A'^2&=& -\frac{2\kappa_g}{f_T}\Big(\Lambda+\rho+\rho_{TB} \Big),\label{0.00006}
\end{eqnarray}
where
\begin{eqnarray}
\kappa_g P_{TB}&=&\frac{1}{4}\Big[f+8(A''+4A'^2)f_B\Big],\\
\kappa_g \rho_{TB}&=& -12\Big[A'(A'''+8A'A'')(f_{BB}+f_{TB})+3A''A'^2(f_{TT}+f_{BT})\Big]\nonumber\\& &+2(A''+4A'^2)f_{B}+\frac{1}{4}f.
\end{eqnarray}
Note that the left side of equations (\ref{0.00005}) and (\ref{0.00006}) is equivalent to that obtained in TERG. So we can states that modified gravity equations of the motion of the $f(T,B)$ gravities is similar to an inclusion of an additional source with $\rho_{TB}$ and $P_{TB}$. 

Once we studied the effects of the torsion on the region exterior to the brane, let us turn out attention to the brane core region. In order to do it, we propose a smooth thick warp factor ansatz of the form \cite{Gremm1999}
\begin{eqnarray}
\label{coreA}
e^{2A(y)}=\cosh^{-2p}(\lambda y),
\end{eqnarray}
where the parameters $p$ and $\lambda$ determine, respectively, the amplitude and the width of the source. Now, we have two important issues, namely, the energy conditions and the field solutions. 

\subsubsection{Energy conditions}

 Let us now analysing the profile of the energy density which yields to the Eq. \eqref{coreA}. Here, we will study how the $f(T,B)$ gravity leads to the brane splitting process.
 
 When, $y\rightarrow \pm \infty$  (vacuum), the stress-energy tensor vanishes, $\mathcal{T}_N\ ^M=0$. So for $f_1(T,B)$, we have
\begin{eqnarray}\label{e.005}
\Lambda=-\frac{1}{\kappa_g}\Big\{3(p\lambda)^2-(-4)^{n_1-1}k_1(n_1-1)[8(p\lambda)^2]^{n_1}\Big\}.
\end{eqnarray}
For $f_2(T,B)$, we have
\begin{eqnarray}\label{e.0005}
\Lambda=\frac{1}{\kappa_g}\Big\{(-4)^{n_2-1}k_2(2n_2-1)[3(p\lambda)^2]^{n_2}\Big\}.
\end{eqnarray}
Comparing equation (\ref{e.1.a}) with (\ref{e.005}) and equation (\ref{e.1.b}) with (\ref{e.0005}), we can say that $A(y\rightarrow \pm \infty)\rightarrow -p\lambda|y|$. Thus, the spacetime solution is asymptotically $AdS_5$.

The energy densities for $f_1(T,B)$ are
\begin{eqnarray}
\rho_1(y)&=&-\Lambda-\frac{3}{\kappa_g}\Big[(p\lambda)^2-\frac{1}{2}(1+2p)p\lambda^2\mathrm{sech}^2(\lambda y)\Big]\nonumber\\
& &+\frac{\alpha}{\kappa_g}\Bigg[2^{3n_1-2}(n_1-1)k_1 -\frac{8^{n_1-1}(n_1-1)3n_1k_1(1+4p)\sinh^2(\lambda y)}{\beta^2}\Bigg] ,
\end{eqnarray}
and pressure
\begin{eqnarray}
P_{1}(y)=-\Lambda-\frac{1}{4\kappa_g}\Big[12(p\lambda)^2\tanh^2(\lambda y) -8^{n_1}\alpha^{n_1}k_1(n_1-1)\Big] ,
\end{eqnarray}
where we defined the functions $\alpha\equiv p \lambda^2[\mathrm{sech}^2(\lambda y)-4p\tanh^2(\lambda y)]$, $\beta\equiv1+2[1-\cosh(2\lambda y)]p$, and $\Lambda$ is given in Eq.(\ref{e.005}).

For $n_1=1$, the source exhibits a localized profile satisfying the dominant and strong energy conditions independent of the parameter $ k_1 $. 

\begin{figure}
\begin{center}
\begin{tabular}{ccc}
\includegraphics[height=5cm]{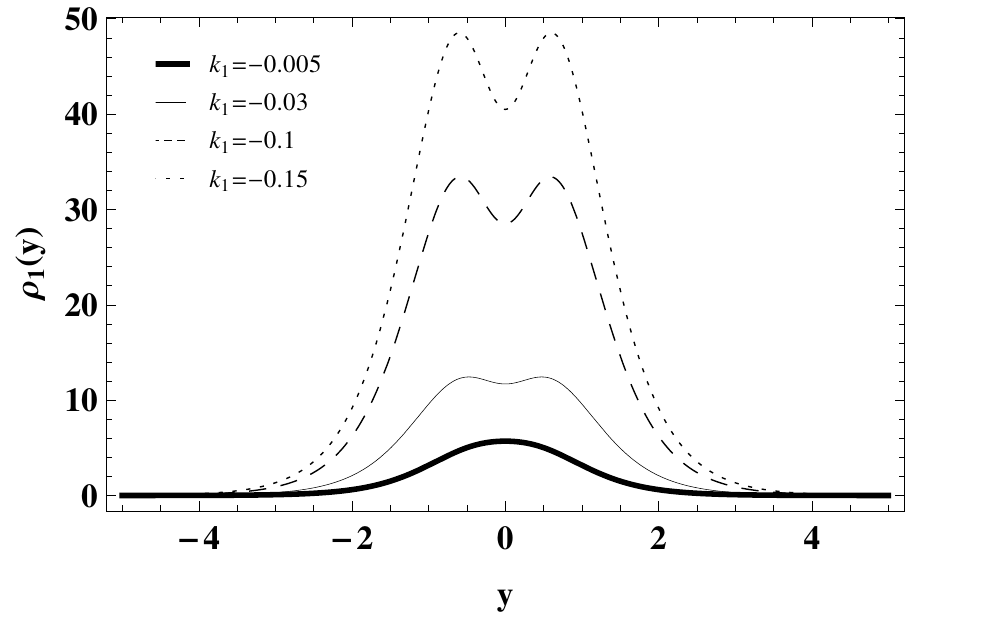}
\includegraphics[height=5cm]{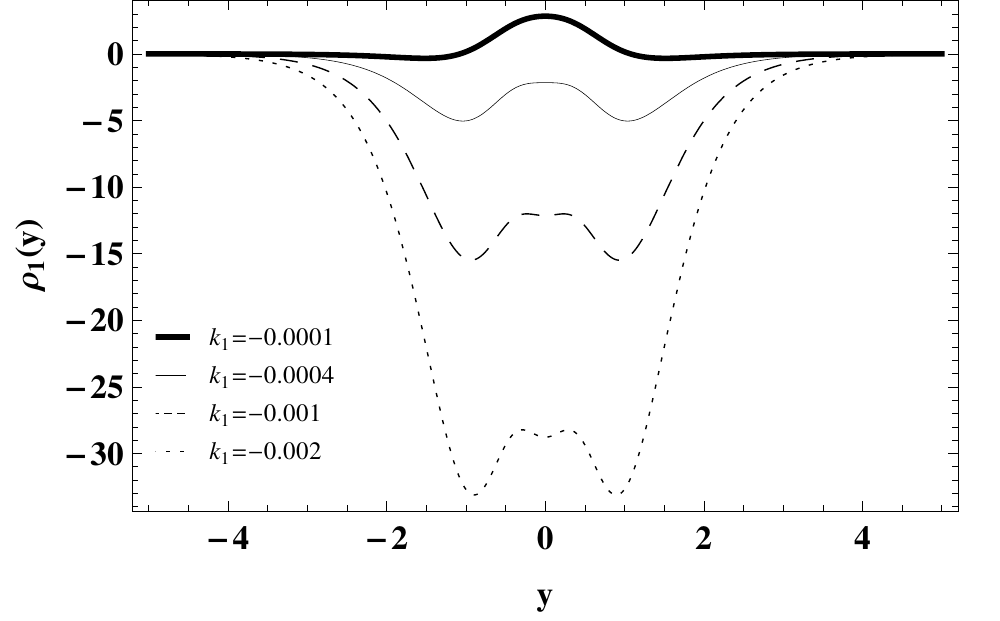}\\ 
(a) \hspace{8 cm}(b)\\
\includegraphics[height=5cm]{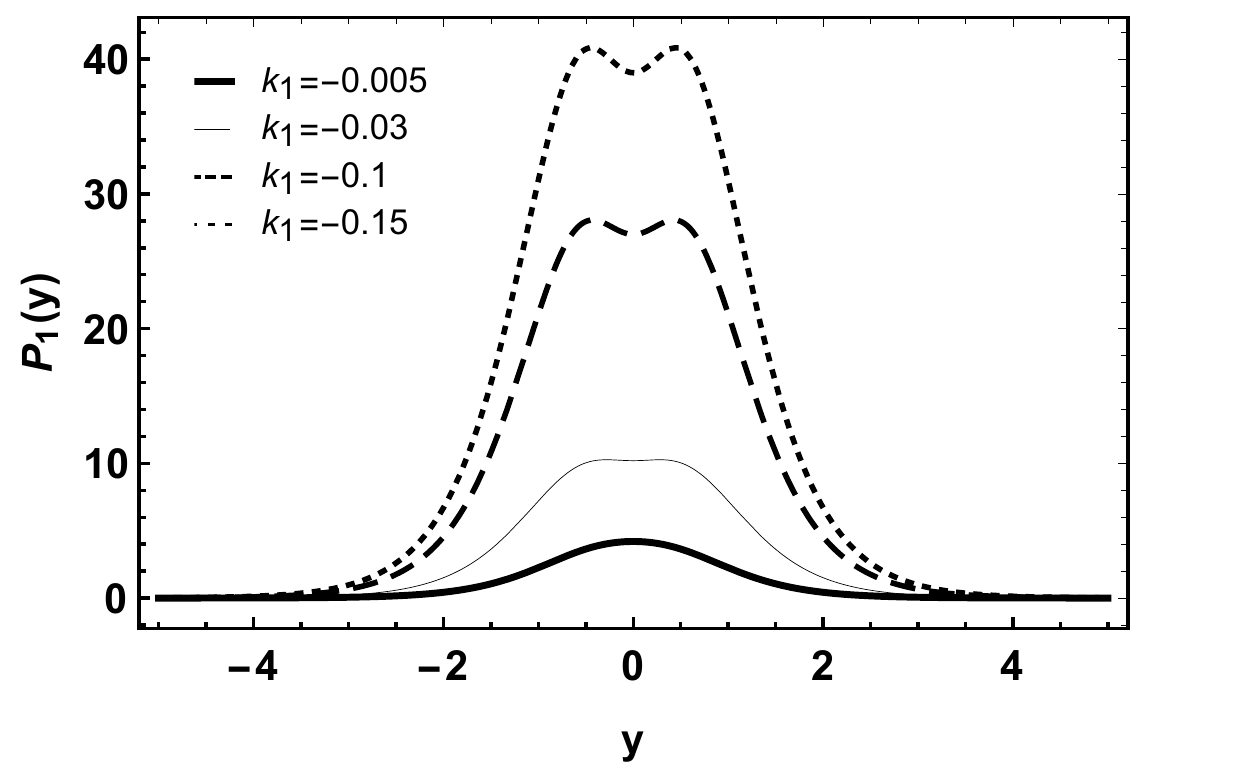} 
\includegraphics[height=5cm]{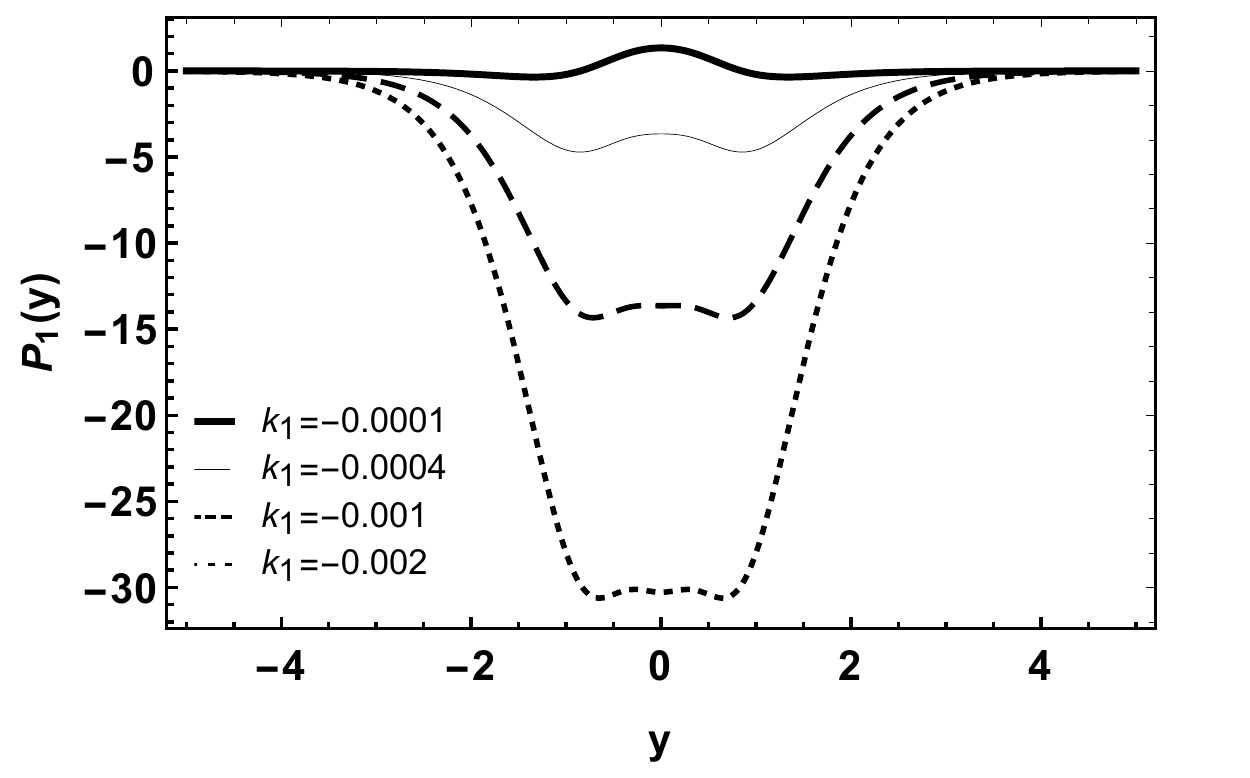}\\
(c) \hspace{8 cm}(d)
\end{tabular}
\end{center}
\caption{Plots of the energy density for $p=\lambda=1$. (a) $n_1=2$ . (b) $n_1=3$. Plots of the pressure $P_{1}(y)$. (c) $n_1=2$ . (d) $n_1=3$.
\label{figene1}}
\end{figure}

In Fig.\ref{figene1}, we plotted the energy densities $\rho_1(y)$  and pressure $P_1(y)$ varying the parameter $k_1$.  
The $n_1=2$ configuration (figure \ref{figene1} $a$ ) includes a new peak. For the $n_1=3$ configuration, we can say that three valleys appear for $k_1=-0.002$ (figure \ref{figene1} $b$). This feature reflects the brane internal structure, which tends to split the brane. A similar splitting process was obtained in Ref. \cite{Yang2012}. 

A noteworthy feature is the violation of the dominant energy condition for $n_1=2$ with $k_1>0$ and $n_1=3$ with $k_1<0$,  where source presents a negative energy density phase.  Therefore, $f_1(T,B)$ produces modifications in the source equation of state that might lead to the brane splitting. For $ n_1 = 3 $ with $ k_1> 0 $, the source has a positive energy density phases.

\begin{figure}
\begin{center}
\begin{tabular}{ccc}
\includegraphics[height=3.5cm]{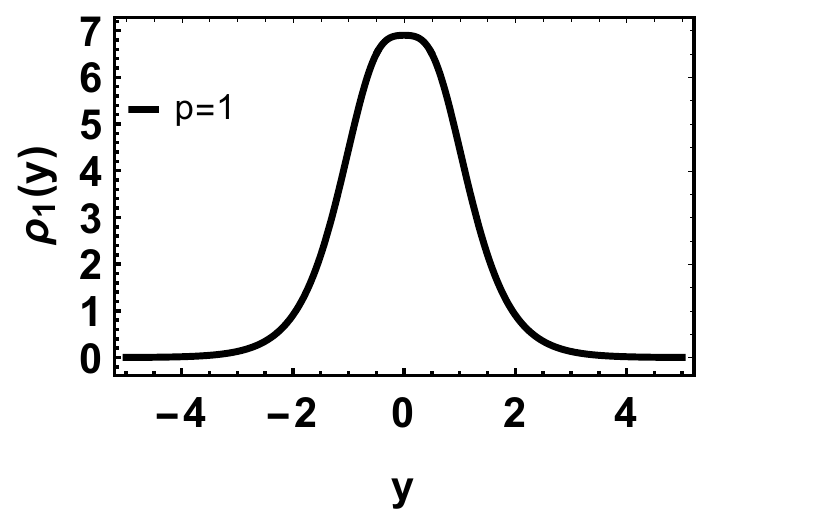}
\includegraphics[height=3.5cm]{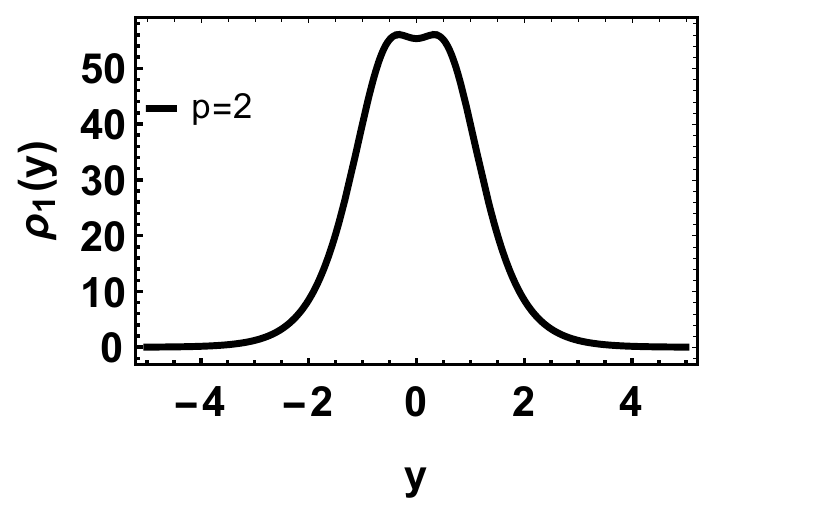}
\includegraphics[height=3.5cm]{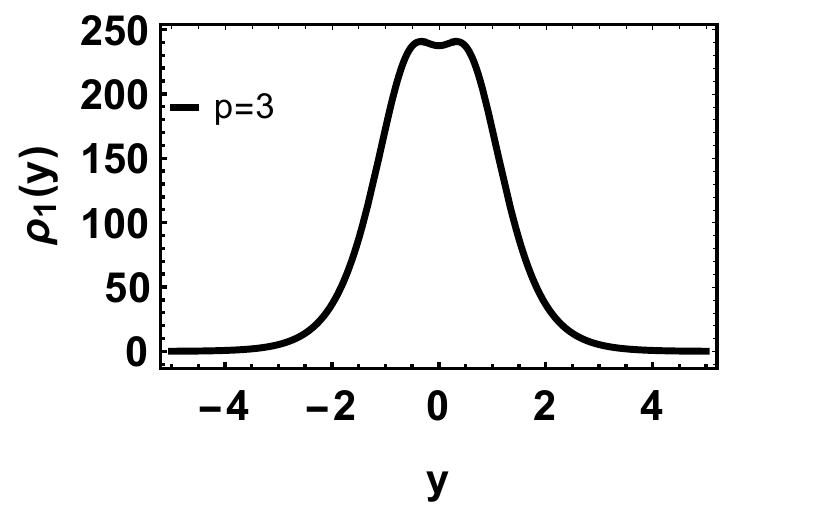}
\end{tabular}
\end{center}
\caption{Plots of the energy density for $n_1=2$ with $\lambda=1$ and $k_1=-0.01$.
\label{figene2}}
\end{figure}

In Fig.\ref{figene2}, we plotted the energy densities $\rho_1(y)$ varying the parameter $p$ for $n_1=2$. Note that by varying the parameter $ p $ we obtain the splitting of the brane. 

The energy densities for $f_2(T,B)$ are
\begin{eqnarray}
\rho_2(y)=-\Lambda-\frac{1}{\kappa_g}\Big\{2^{2n_2-3}(-3)^{n_2}p^{-1}k_2(2n_2-1)\beta\mathrm{csch}^2(\lambda y)[p\lambda\tanh(\lambda y)]^{2n_2}\Big\},
\end{eqnarray}
and pressure
\begin{eqnarray}
P_{2}(y)=-\Lambda-\frac{1}{\kappa_g}\Big\{(-3)^{n_2}4^{n_2-1}k_2(2n_2-1)[(p\lambda)^2\tanh^2(\lambda y)]^{n_2}\Big\} ,
\end{eqnarray}
where $\Lambda$ is given in Eq.(\ref{e.0005}).

For $n_2=1$, the source exhibits a localized profile satisfying the dominant and strong energy conditions, but when $k_2 <0$, the source presents a negative energy density phase. 

\begin{figure}
\begin{center}
\begin{tabular}{ccc}
\includegraphics[height=5cm]{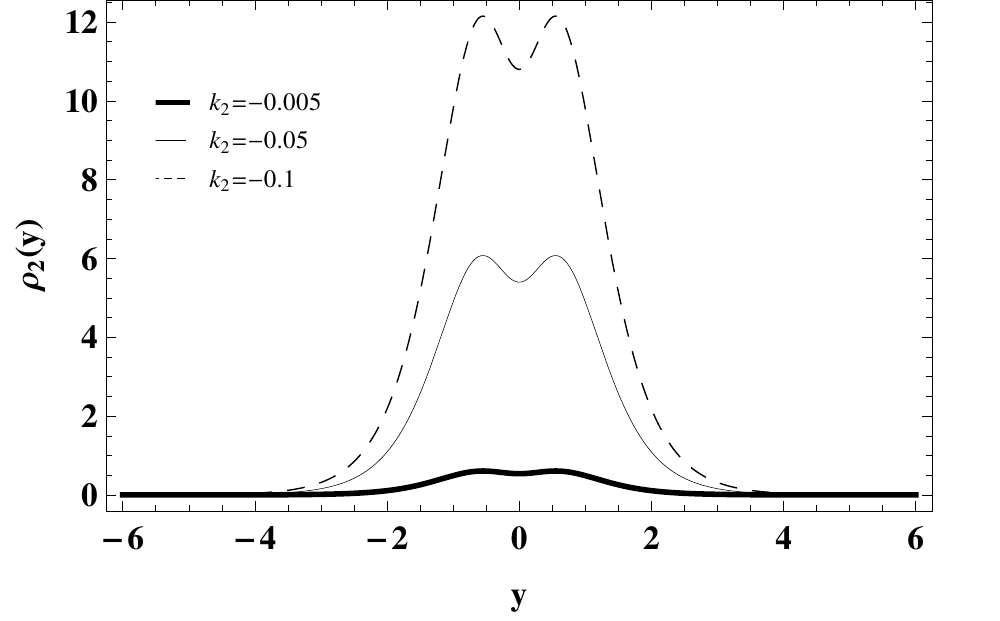}
\includegraphics[height=5cm]{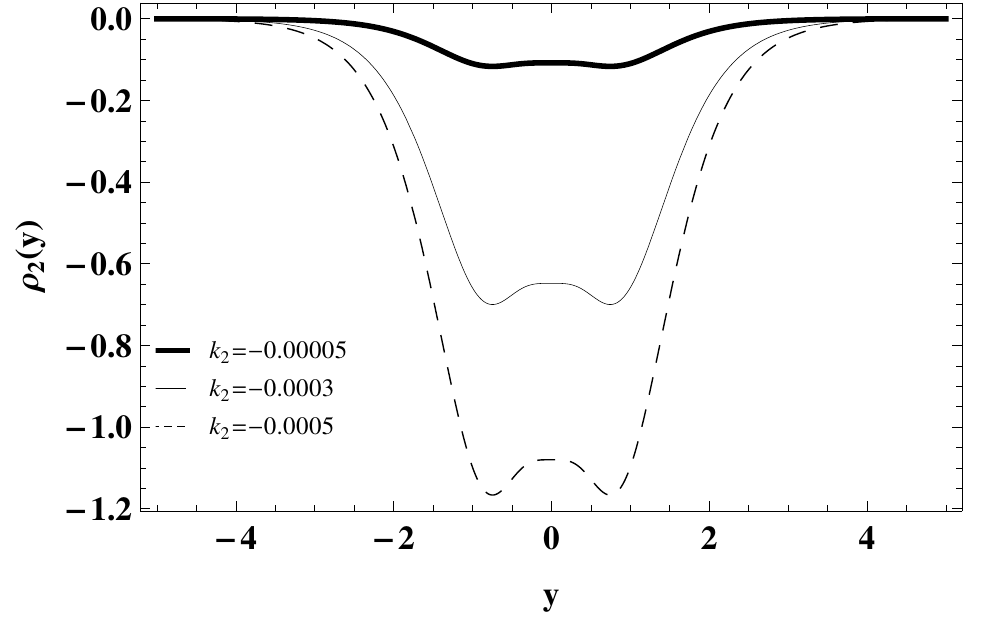}\\ 
(a) \hspace{8 cm}(b)\\
\includegraphics[height=5cm]{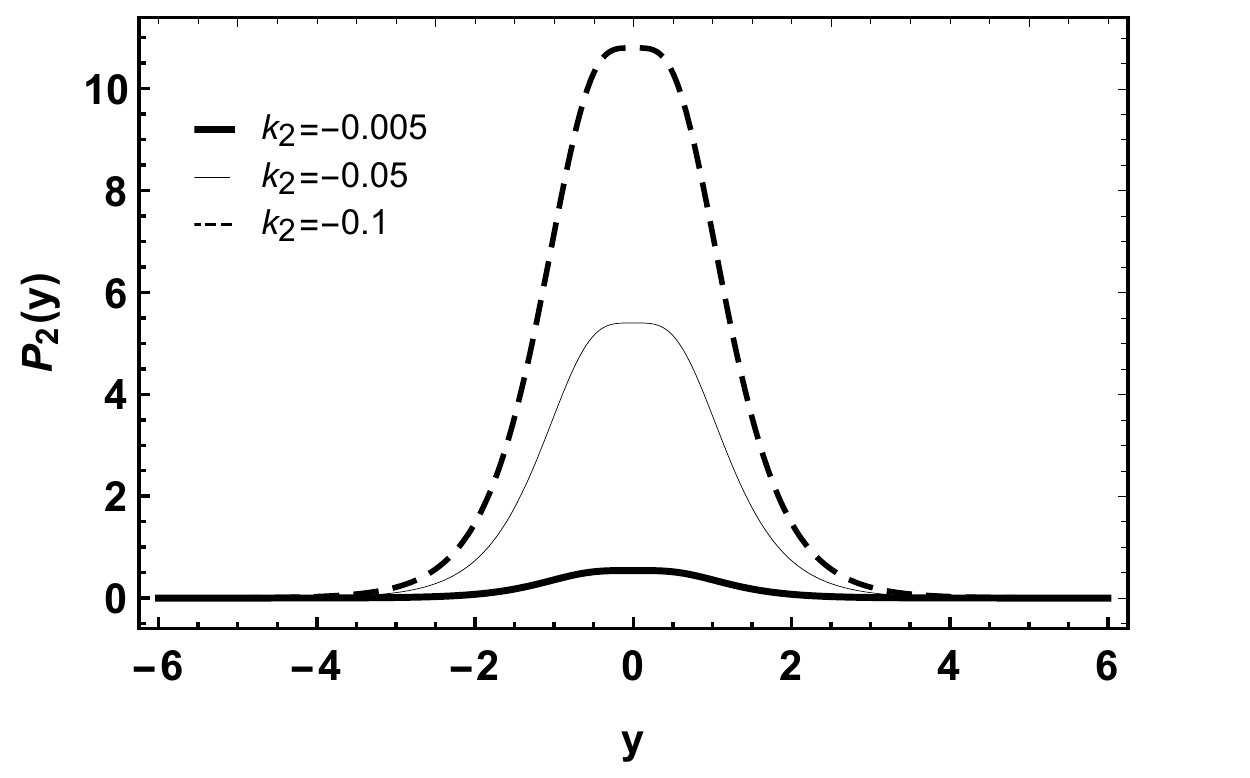} 
\includegraphics[height=5cm]{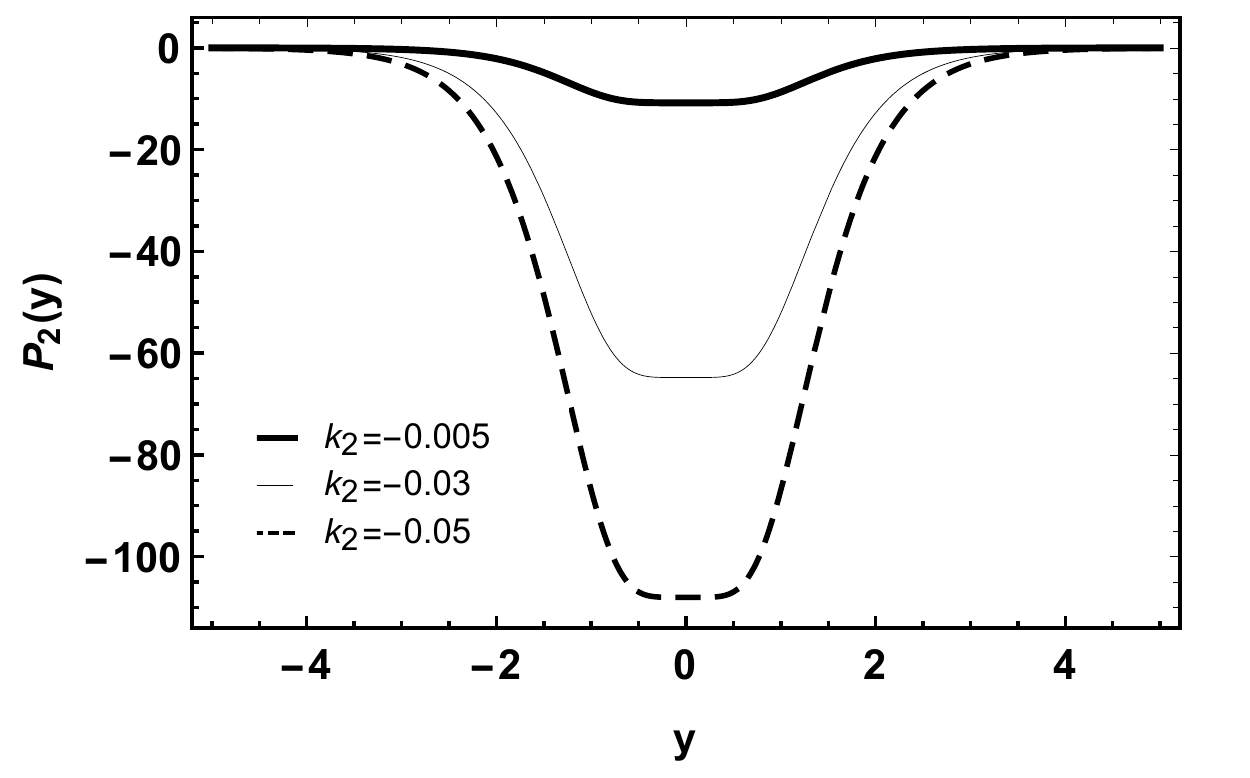}\\
(c) \hspace{8 cm}(d)
\end{tabular}
\end{center}
\caption{Plots of the energy density for $p=\lambda=1$. (a) $n_2=2$ . (b) $n_2=3$. Plots of the pressure $P_{2}(y)$. (c) $n_2=2$ . (d) $n_2=3$.
\label{figene3}}
\end{figure}

In Fig.\ref{figene3}, we plotted the energy densities $\rho_2(y)$ and pressure $P_2(y)$ varying the parameter $k_2$. For $n_2=2$ and $n_2=3$ configuration, the energy densities   includes a new peak regardless of the $k_2$ parameter.  Again, this feature reflects the brane internal structure, which tends to split the brane.  

As in $f_1(T,B)$, we noticed the violation of the dominant energy condition for $n_2=2$ with $k_2>0$ and $n_2=3$ with $k_2<0$ , where source presents a negative energy density phase.  Therefore, the $f_2(T,B)$ produces modifications in the source equation of state that might lead to the brane splitting. For $ n_2 = 3 $ with $ k_2> 0 $, the source has a positive energy density phases.

\begin{figure}
\begin{center}
\begin{tabular}{ccc}
\includegraphics[height=3.5cm]{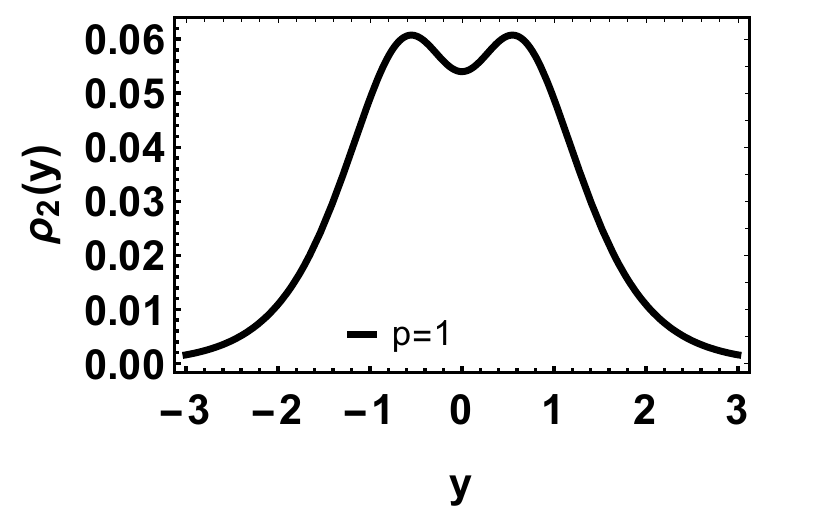}
\includegraphics[height=3.5cm]{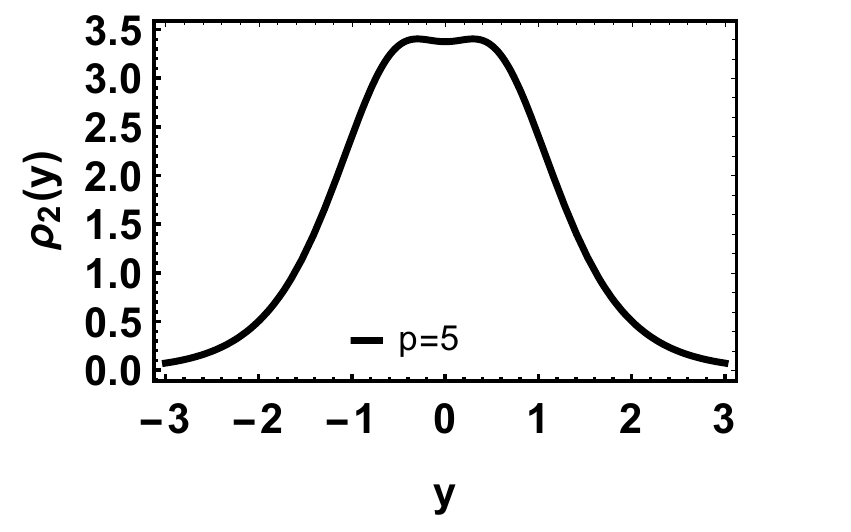}
\includegraphics[height=3.5cm]{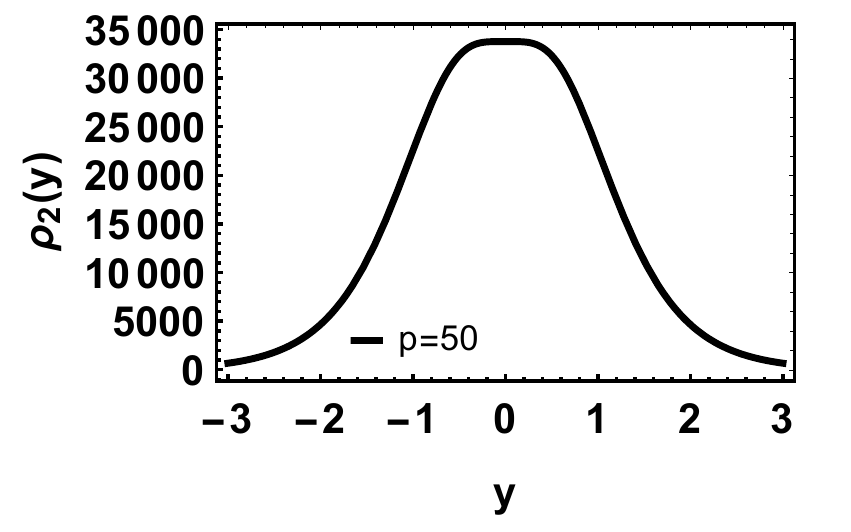}
\end{tabular}
\end{center}
\caption{Plots of the energy density for $n_2=2$ with $\lambda=1$ and $k_2=-0.00005$.
\label{figene4}}
\end{figure}

In Fig.\ref{figene4}, we plotted the energy densities $\rho_2(y)$ varying the parameter $p$ for $n_2=2$. Note that the higher the value of $p$, we undo the splitting in the brane. 

\subsubsection{Field solution}

In this subsection, we obtain the configurations of the scalar field which leads to the thick brane \eqref{coreA}.

We follow the approach carried out in Ref.\cite{Yang2012}, where by manipulating the modified Einstein equations, an equation relating the metric components and the scalar field was obtained. In this case we can write Eq. (\ref{e.1}) and Eq. (\ref{e.2}) as 
\begin{eqnarray}\label{q.4}
\phi'^2(y)=-\frac{3}{2}f_TA''+12A'\Big[(8A'A''+A''')(f_{BB}+f_{TB})+3A'A''(f_{TT}+f_{BT})\Big],
\end{eqnarray}
\begin{eqnarray}\label{q.44}
V(\phi)&=&6A'\Big[(8A'A''+A''')(f_{BB}+f_{TB})+3A'A''(f_{TT}+f_{BT})\Big]\nonumber\\&-&\frac{1}{4}\Big[3(8A'^2+A'')f_T+8(4A'^2+A'')f_B+f\Big],
\end{eqnarray}
where we set the gravitational constant $\kappa_g=1$ for simplicity.   The Eq.(\ref{q.4}) allows us to obtain the scalar field for a given geometric solution \cite{Yang2012}. As one knows, solutions $\phi= \phi(y)$ of the above equation (\ref{q.4}) must go to some constant value $\phi_c$ asymptotically. Thus, if we want that our model makes physical sense, the potential (\ref{q.44}) should go to a vacuum value for $\phi_c$.

For $f_1(T,B)$ the Eqs.(\ref{q.4}) and (\ref{q.44}) takes the form
\begin{eqnarray}\label{q.5}
\phi'^2(y)=\frac{3}{2}p\lambda^2\mathrm{sech}^2(\lambda y)-\frac{\alpha^{n_1}}{\beta^2}\Big[8^{n_1-1}k_1n_1(n_1-1)(1+4p)\sinh^2(\lambda y)\Big],
\end{eqnarray}
\begin{eqnarray}\label{q.55}
V(\phi(y))&=&\frac{3}{4}\alpha+2^{3n_1-4}k_1(n_1-1)(p\lambda^2)^2\alpha^{n_1-2}\Big\{4 \mathrm{sech}^4(\lambda y)+64p^2\tanh ^4(\lambda y)\nonumber\\ &-&[3n_1+4(8+3n_1)]\mathrm{sech}^2(\lambda y)\tanh ^2(\lambda y)\Big\}.
\end{eqnarray}
In this case, with equation (\ref{e.005}), the asymptotic values of the potential and its derivative with respect to the field are, respectively,
\begin{eqnarray}
\Lambda\equiv V(\phi\rightarrow\pm\phi_{c_1})=(-4)^{n_1-1}k_1(n_1-1)[8(p\lambda)^2]^{n_1}-3(p\lambda)^2,
\end{eqnarray}
and $\partial V(\phi\rightarrow\pm\phi_{c_1})/\partial \phi=0$.
We can solve Eq. (\ref{q.5}) to find a function $\phi=g(y)$ that may be inverted to give $y=g^{-1}(\phi)$, which allows us to write the potential in the usual way $V=V(\phi)$. The thick brane solution for $n_1=1$ and the potential are
\begin{eqnarray}
\phi(y)=\sqrt{6p}\ \mathrm{arctan}\Big[\tanh\Big(\frac{\lambda y}{2}\Big)\Big],
\end{eqnarray}
\begin{eqnarray}
V(\phi)=\frac{3p\lambda^2}{4}\Big[(1+4p)\cos^2\Big(\frac{2\phi}{\sqrt{6p}}\Big)-4p\Big],
\end{eqnarray}
that are the same expressions as those in Refs.\cite{Gremm1999,Bazeia2007}. One point here is that the expressions there were obtained by using a superpotential approach where the gravity is described by general relativity. As a consequence, the solutions of this case are equivalent to those in the case $f_1(T,B)=T$. For $n_1=2$
\begin{eqnarray}\label{q.66}
\phi(y)&=&\sqrt{3p}\lambda\coth(\lambda y)\sqrt{1-32[1-\mathrm{sech}^2(\lambda y)]}\Bigg\{\frac{1}{\varrho}\Big[\tau \mathrm{E}\Big(\mathrm{arcsin}[\cosh(\lambda y)], 1-\frac{1}{\tau}\Big)\nonumber\\ &-& \mathrm{F}\Big(\mathrm{arcsin}[\cosh(\lambda y)], 1-\frac{1}{\tau}\Big)\Big]+\frac{\mathrm{sech}(\lambda y)}{[1+\cos(2\lambda y)}]\Bigg\},
\end{eqnarray}
where $\tau=32k_1p(1+4p)\lambda^2$ and $\varrho=2\sqrt{2[1+\cosh(2\lambda y)]}\sqrt{(\tau-1)\cosh(2\lambda y)-\tau-1}$, and $\mathrm{F}(y,q)$, $\mathrm{E}(y,q)$ are the first and second kind elliptic integrals, respectively. Note that $k_1\geq 1/32p(1+4p)\lambda^2$ for the second solution (\ref{q.66})  in order to ensure that the scalar field $\phi(y)$ is real. In this case, the $V(\phi)$ solution cannot be found analytically.

For $f_2(T,B)$  Eqs.(\ref{q.4} and \ref{q.44}) takes the form
\begin{eqnarray}\label{q.6}
\phi'^2(y)=-2^{2n_2-3}(-3)^{n_2}p^{-1}k_2n_2(2n_2-1)\mathrm{csch}^2(\lambda y)[p\lambda\tanh(\lambda y)]^{2n_2},
\end{eqnarray}
\begin{eqnarray}\label{q.606}
V(\phi(y))&=&4^{n_2-2}(-3)^{n_2}p^{-1}k_2(2n_2-1)[4p-n_2\mathrm{csch}^2(\lambda y)][p\lambda\tanh (\lambda y)]^{2n_2}.
\end{eqnarray}
In this case, with equation (\ref{e.0005}), the asymptotic value of the potential is 
\begin{eqnarray}
\Lambda\equiv V(\phi\rightarrow\pm\phi_{c_2})=(-4)^{n_2-1}k_2(2n_2-1)[3(p\lambda)^2]^{n_2},
\end{eqnarray}
and its derivative in relation to the field is $\partial V(\phi\rightarrow\pm\phi_{c_2})/\partial \phi=0$.
The thick brane solution for $n_2=1$ and the potential are
\begin{eqnarray}
\phi(y)=\sqrt{6pk_2}\ \mathrm{arctan}\Big[\tanh\Big(\frac{\lambda y}{2}\Big)\Big],
\end{eqnarray}
\begin{eqnarray}
V(\phi)=\frac{3pk_2\lambda^2}{4}\Big[(1+4pk_2)\cos^2\Big(\frac{2\phi}{\sqrt{6pk_2}}\Big)-4pk_2\Big].
\end{eqnarray}
For case $n_2=2$, we have
\begin{eqnarray}
\phi(y)=-6\sqrt{-3p^3k_2}\lambda\ \mathrm{sech}(\lambda y),
\end{eqnarray}
\begin{eqnarray}
V(\phi)=108k_2(p\lambda)^4+\frac{\lambda^2}{2}(1+4p)\phi^2+\frac{1+2p}{216k_2p^3}\phi^4,
\end{eqnarray}
which is not a pleasant solution. The same goes for cases where $ n_2 = 4, 6,... $ (even numbers). When $n_2=3$, we have
\begin{eqnarray}
\phi(y)=9(p\lambda)^2\sqrt{10pk_2}\{ \mathrm{arctan}[\sinh(\lambda y)]-\mathrm{sech}(\lambda y)\tanh(\lambda y)\}.
\end{eqnarray}
In this case the $V(\phi)$ solution cannot be found analytically.
We note that for odd $n_2$, the parameter $k_2$ must be greater than zero to obtain real solutions of $\phi(y)$. For even $n_2$, the parameter must be negative ($k_2<0$) in order to obtain real solutions of $\phi(y)$.

\section{Gravitational Tensor Modes}
\label{sec2}

In this section we investigate the effects of torsion and boundary term on the propagation of the linear perturbations in the brane system. We follow closely the analysis performed in Ref.\cite{tensorperturbations}. For that, we make the \textit{f\"unfbein} perturbation 
\begin{eqnarray}
h^a\ _M=\left(\begin{array}{cccccc}
e^{A(y)}\left(\delta^a_\mu+w^a\ _\mu\right)&0\\
0&1\\
\end{array}\right),
\end{eqnarray}
where $w^a\ _\mu=w^a\ _\mu(x^\mu,y)$. The resulting metric perturbation takes the form $ds^2=e^{A(y)}\left(\eta_{\mu\nu}+\gamma_{\mu\nu}\right)dx^\mu dx^\nu+dy^2$,
where the metric and the \textit{f\"unfbein} perturbations are related by
\begin{eqnarray}
 \gamma_{\mu\nu}&=&(\delta^a_\mu w^b\ _\nu+\delta^b_\nu w^a\  _\mu)\eta_{ab},\nonumber\\
 \gamma^{\mu\nu}&=&(\delta_a^\mu w_b\ ^\nu+\delta_b^\nu w_a\  ^\mu)\eta^{ab}.
\end{eqnarray}
We assume the transverse-traceless (TT) tensor perturbation, which is related to the gravitational wave and
four-dimensional gravitons. The TT tensor perturbation satisfies the following TT conditions $\partial_\mu \gamma^{\mu\nu}=0=\eta^{\mu\nu}\gamma_{\mu\nu}$, which leads to the \textit{f\"unfbein} 
\begin{equation}
\partial_\mu\left(\delta_a^\mu w_b\ ^\nu+\delta_b^\nu w_a\  ^\mu\right)\eta^{ab}=0,
\end{equation}
\begin{equation}\label{06665}
\delta_a^\mu w^a\  _\mu=0.
\end{equation}
The non-vanishing components of the torsion tensor are \cite{tensorperturbations}
\begin{eqnarray}
\label{23.l}
 T^\rho\ _{\mu y}&=&-A'\delta^\rho_\mu-(\delta^\rho_a w^a\ _\mu-\delta^a_\mu w_a\ ^\rho)A'-\delta^\rho_a w^a\ _\mu, \nonumber \\
 T^\rho\ _{\mu \nu}&=&\delta^\rho_a(\partial_\mu w^a\ _\nu-\partial_\nu w^a\ _\mu),
\end{eqnarray}
whereas the non-vanishing contorsion components are \cite{tensorperturbations}
\begin{eqnarray}\label{24.l}
 K^\rho\ _{\mu y}&=&A'(\delta^\rho_a w^a\ _\mu-\delta^a_\mu w_a\ ^\rho)+\frac{1}{2}\left(\delta^\rho_a w'^a\ _\mu-\delta^a_\mu w'_a\ ^\rho\right),\nonumber\\
 K^\rho\ _{y \nu}&=&-A'\delta^\rho_\nu-\frac{1}{2}\left(\delta^a_\nu w'_a\ ^\rho-\delta_a^\rho w'^a\ _\nu\right),\nonumber\\
 K^y\ _{\mu \nu}&=&e^{2A}(A'\eta_{\mu\nu}+A'\gamma_{\mu\nu}+\frac{1}{2}\gamma'_{\mu\nu}),\nonumber\\
 K^\rho\ _{\mu \nu}&=& \frac{1}{2}\left[\delta^a_\mu(\partial^\rho w_{a\nu}-\partial_\nu w_a\ ^\rho)+\delta^a_\nu(\partial^\rho w_{a\mu}-\partial_\mu w_a\ ^\rho)-\delta_a^\rho(\partial_\mu w^a\ _{\nu}-\partial_\nu w^a\ _\mu)\right].
\end{eqnarray}
Accordingly, the non-vanishing components of the dual torsion tensor are \cite{tensorperturbations}
\begin{eqnarray}\label{25.l}
 S_\rho\ ^{\mu y}&=&\frac{1}{2}\left[(3A'+B')\delta^\mu_\rho-\frac{1}{2}(\delta_\rho^a w'_a\ ^\mu+\delta_a^\mu w'^a\ _\rho)\right],\nonumber\\
 S_y\ ^{\mu \nu}&=&\frac{1}{2}\left[A'(\delta^\mu_a w^{a\nu}-\delta_a^\nu w^{a\mu})+\frac{1}{2}(\delta^\mu_a w'^{a\nu}-\delta_a^\nu w'^{a\mu})\right]e^{-2A},\nonumber\\
 S_\rho\ ^{\mu \nu}&=&\frac{1}{4}\left[\delta^\nu_a(\partial^\mu w^{a}\ _\rho-\partial_\rho w^{a\mu})-\delta^\mu_a(\partial^\nu w^{a}\ _\rho-\partial_\rho w^{a\nu})\right]e^{-2A}+\frac{1}{4}\delta_\mu^a(\partial^\mu w_{a}\ ^\nu-\partial^\nu w_a\ ^{\mu})\nonumber\\ &+&
\frac{1}{2}\left[\delta^\nu_\rho\delta^\lambda_a\partial_\lambda w^{a\mu}-\delta^\mu_\rho\delta^\lambda_a\partial_\lambda w^{a\nu}\right]e^{-2A},\nonumber\\
 S_y\ ^{\mu y}&=&\frac{1}{2}(\delta^\rho_a\partial_\rho w^{a\mu})e^{-2A}.
\end{eqnarray}
In this work, we always neglect the second-order terms for the disturbed quantities. Considering the traceless condition (\ref{06665}), we have \cite{tensorperturbations}
\begin{equation}
\delta h= h h_a\ ^M\delta h^a\ _M=h e^{-A}e^A\delta^\mu_aw^a\ _\mu=0.
\end{equation}
After a lenghty but simple algebra it is straightforward to verify that $\delta T=0$ and $\delta B=0$ \cite{tensorperturbations,ftnoncanonicalscalar,ftborninfeld,ftmimetic}. 
 The perturbed modified Einstein equation (\ref{3.36}) has now the form
\begin{eqnarray}\label{27.l}
\frac{1}{4}(Bf_B-f)\delta g_{MN}-\frac{1}{h}f_T\Big[\delta g_{PN}\partial_Q(h S_M\ ^{PQ})+ g_{PN}\partial_Q(h \delta S_M\ ^{PQ})&\nonumber\\
-h\Big(\delta\widetilde{\Gamma}^Q\ _{PM}S_{QN}\ ^{P}+\widetilde{\Gamma}^Q\ _{PM}\delta S_{QN}\ ^{P}\Big)\Big] -\Big[(\partial_Qf_B)+(\partial_Qf_T)\Big]\delta S_{MN}\ ^{Q}\nonumber\\
+\frac{1}{2}(\delta g_{MN}\Box f_B-\delta g_{PN}\nabla^P\nabla_Mf_B)&=&\delta\mathcal{T}_{MN},
\end{eqnarray}
yields to
\begin{eqnarray}\label{28.l}
\frac{1}{4}\Big\{\Big[e^{-2A}\Box^{(4)} \gamma_{\mu\nu}+4A'\gamma'_{\mu\nu}+\gamma''_{\mu\nu}-6(A''+4A'^2)\gamma_{\mu\nu}\Big]f_T& &\nonumber\\
+\Big[8(A'''+8A'A'')(f_{BB}+f_{TB})+24A'A''(f_{TT}+f_{BT})\Big](6A'\gamma_{\mu\nu}-\gamma'_{\mu\nu})& &\nonumber\\
-[f+8(A''+4A'^2)f_B]\gamma_{\mu\nu}\Big\}e^{2A}&=&\delta\mathcal{T}_{\mu\nu},
\end{eqnarray}
where $\Box^{(4)}=\eta^{\mu\nu}\partial_\mu \partial_\nu$. Here, we note that the perturbations vanish in the extra dimension. The linearlized stress energy tensor writes 
\begin{eqnarray}\label{29.l}
\delta\mathcal{T}_{\mu\nu}=\delta(\mathcal{T}_{\mu}\ ^\mu g_{\mu\nu})=\delta(\mathcal{T}_{\mu}\ ^\mu)\eta_{\mu\nu}e^{2A}+\mathcal{T}_{\mu}\ ^\mu \gamma_{\mu\nu}e^{2A}.
\end{eqnarray}
The gravitational field equation (\ref{3.36})  provides the condition
\begin{eqnarray}\label{30.l}
12\Big[A'(A'''+8A'A'')(f_{BB}+f_{TB})+3A''A'^2(f_{TT}+f_{BT})\Big]& &\nonumber\\
-\frac{1}{2}(A''+4A'^2)(4f_{B}+3f_{T})-\frac{1}{4}f&=& \mathcal{T}_{\mu}\  ^\mu.
\end{eqnarray}
By plugging Eq.(\ref{30.l}) into Eq.(\ref{28.l}), and considering  the vanishing trace $\delta(\mathcal{T}_{\mu}\ ^\mu)$, we obtain the following perturbation equation
\begin{eqnarray}\label{32.l}
 -8 \Big[(A'''+8A'A'')(f_{BB}+f_{TB})+3A'A''(f_{TT}+f_{BT})\Big]\gamma'_{\mu\nu}& &\nonumber\\ 
+(e^{-2A}\Box^{(4)} \gamma_{\mu\nu}+4A'\gamma'_{\mu\nu}+\gamma''_{\mu\nu})f_T&=&0.
\end{eqnarray}

Assuming the Kaluza-Klein decomposition $\gamma_{\mu\nu}(x^\rho,y)=\epsilon_{\mu\nu}(x^\rho)\chi(y)$ and a 4D plane-wave satisfying  $\left(\Box^{(4)}-m^2\right)\epsilon_{\mu\nu}=0$, the perturbed Einstein equation (Eq.(\ref{32.l})) yields to
\begin{eqnarray}
\label{KKequation}
\chi''+\Big\{4A'-\frac{1}{f_T} \Big[24A'A''(f_{TT}+f_{TB})+8(A'''+8A'A'')(f_{BB}+f_{TB})\Big]\Big\}\chi'& &\nonumber\\
+e^{-2A}m^2\chi&=&0.
\end{eqnarray}
\subsection{Massive modes}

Let us firstly consider the effects of torsion in the region exterior to the brane. That limit can also be interpreted as representing a thin brane. In this regime $A'=-c$ and the Eq.(\ref{KKequation}) takes the form
\begin{eqnarray}\label{91.l}
\chi''-4c\chi'+e^{2c|y|}m^2\chi=0.
\end{eqnarray}
The Eq.(\ref{91.l}) is the same of the thin brane in RS model \cite{rs}, except that the coefficient of the the first-derivative term is 2 instead of 4 and that $c$ depends on the parameters that control the torsion and  boundary term $n_{1,2}$ and $k_{1,2}$.

\begin{figure}
\begin{center}
\begin{tabular}{ccccccccc}
\includegraphics[height=6cm]{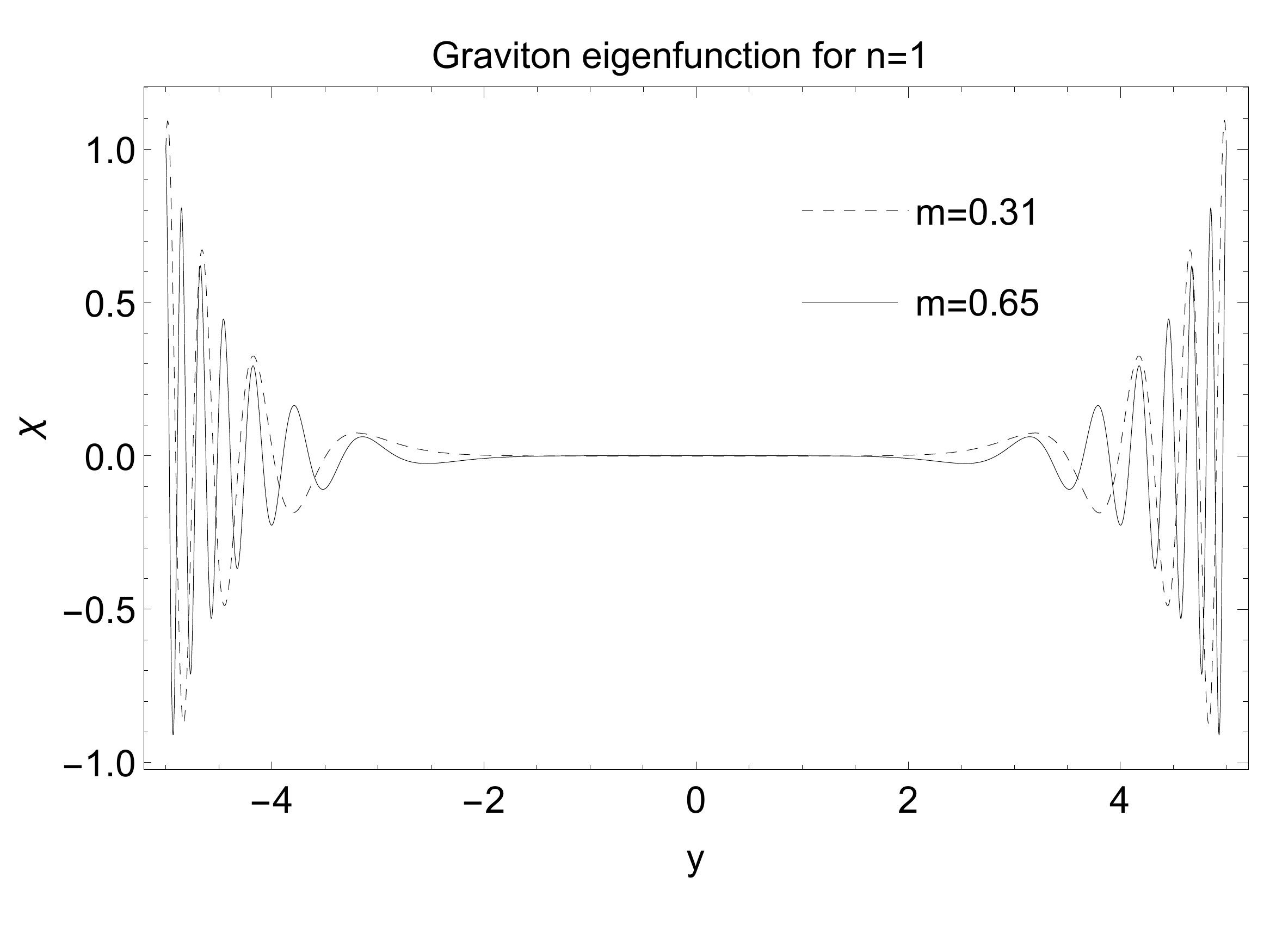}
\includegraphics[height=6cm]{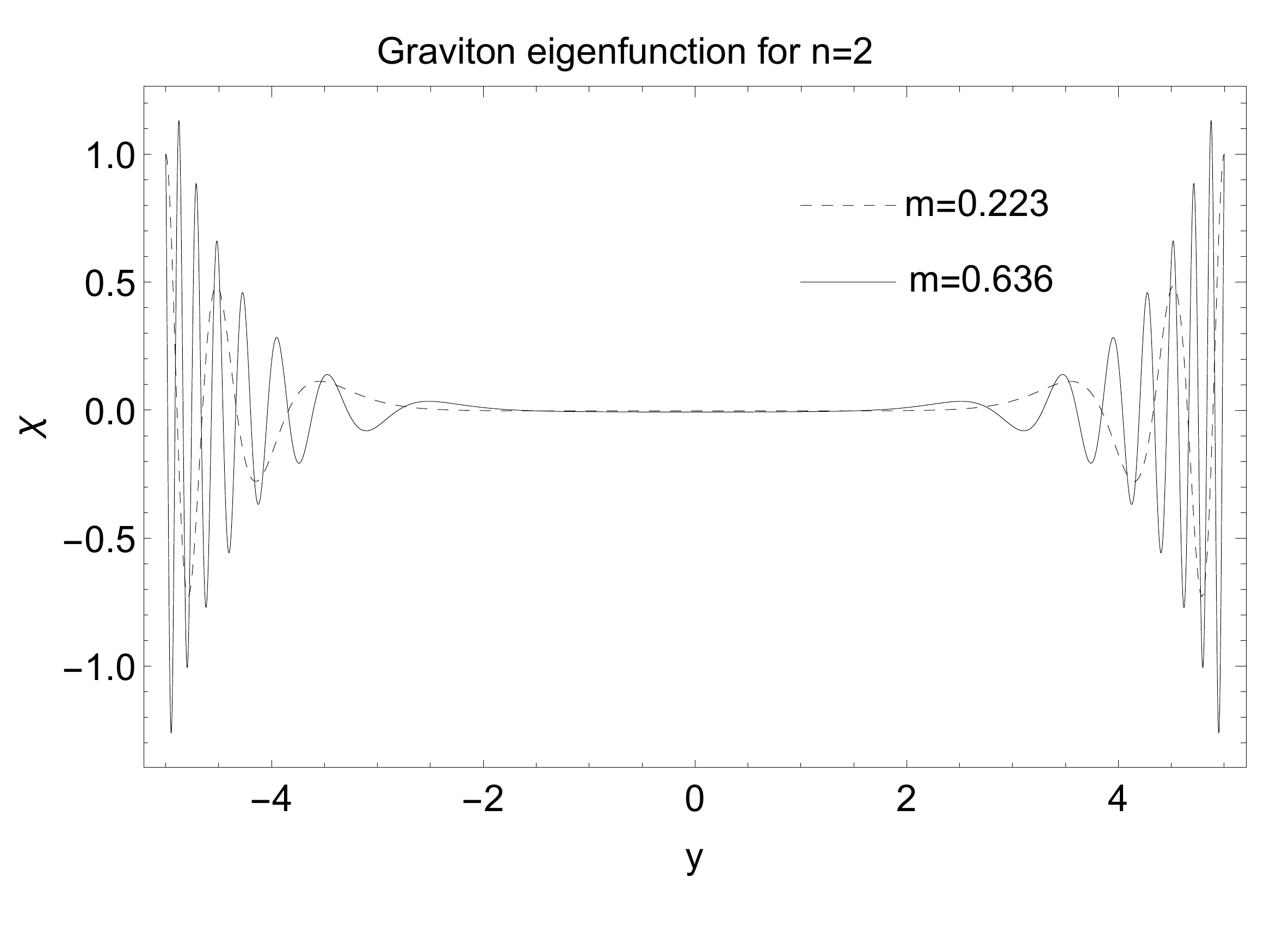}  \\
(a)\hspace{8 cm}(b)\\
\includegraphics[height=6cm]{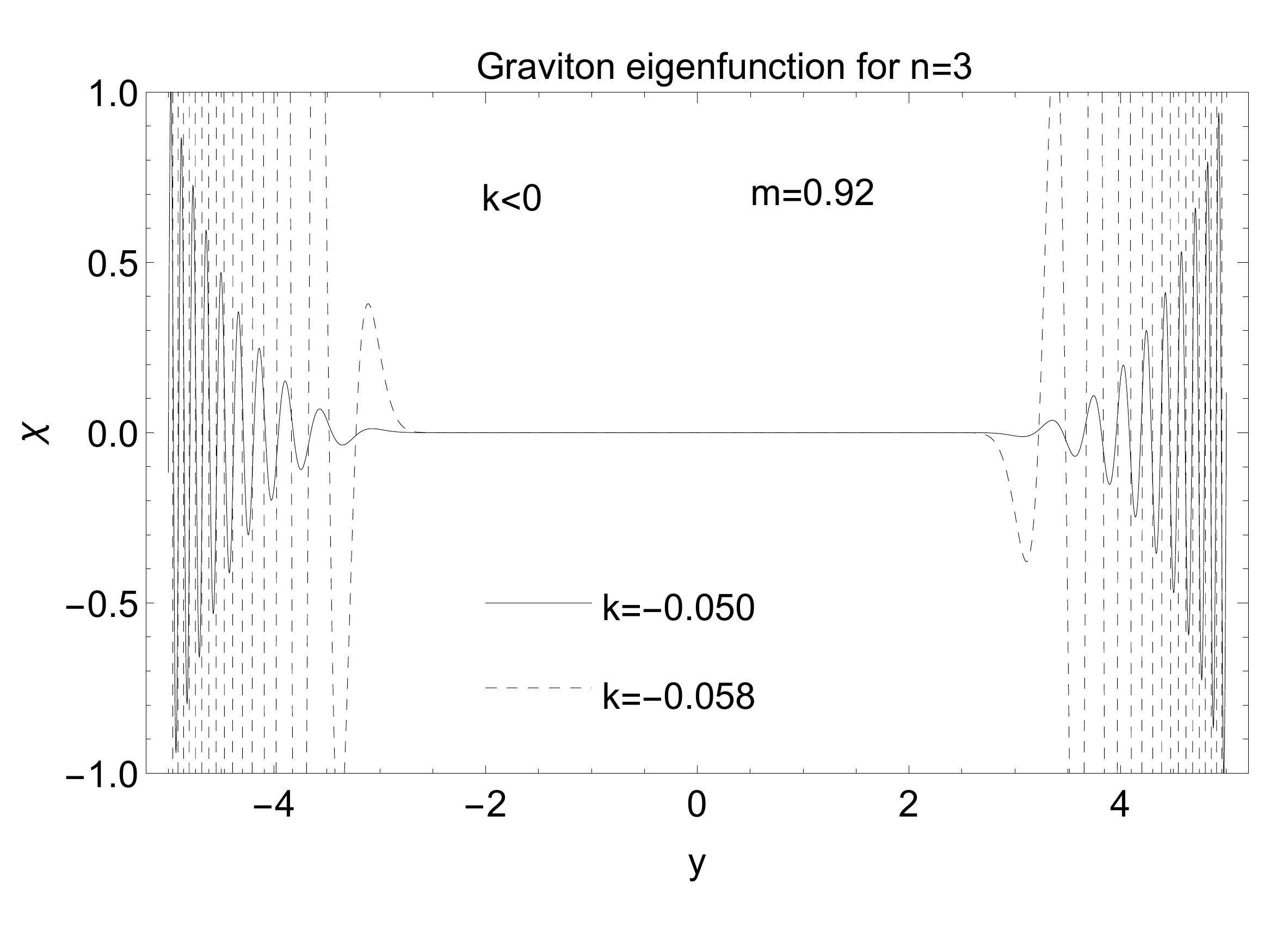} 
\includegraphics[height=6cm]{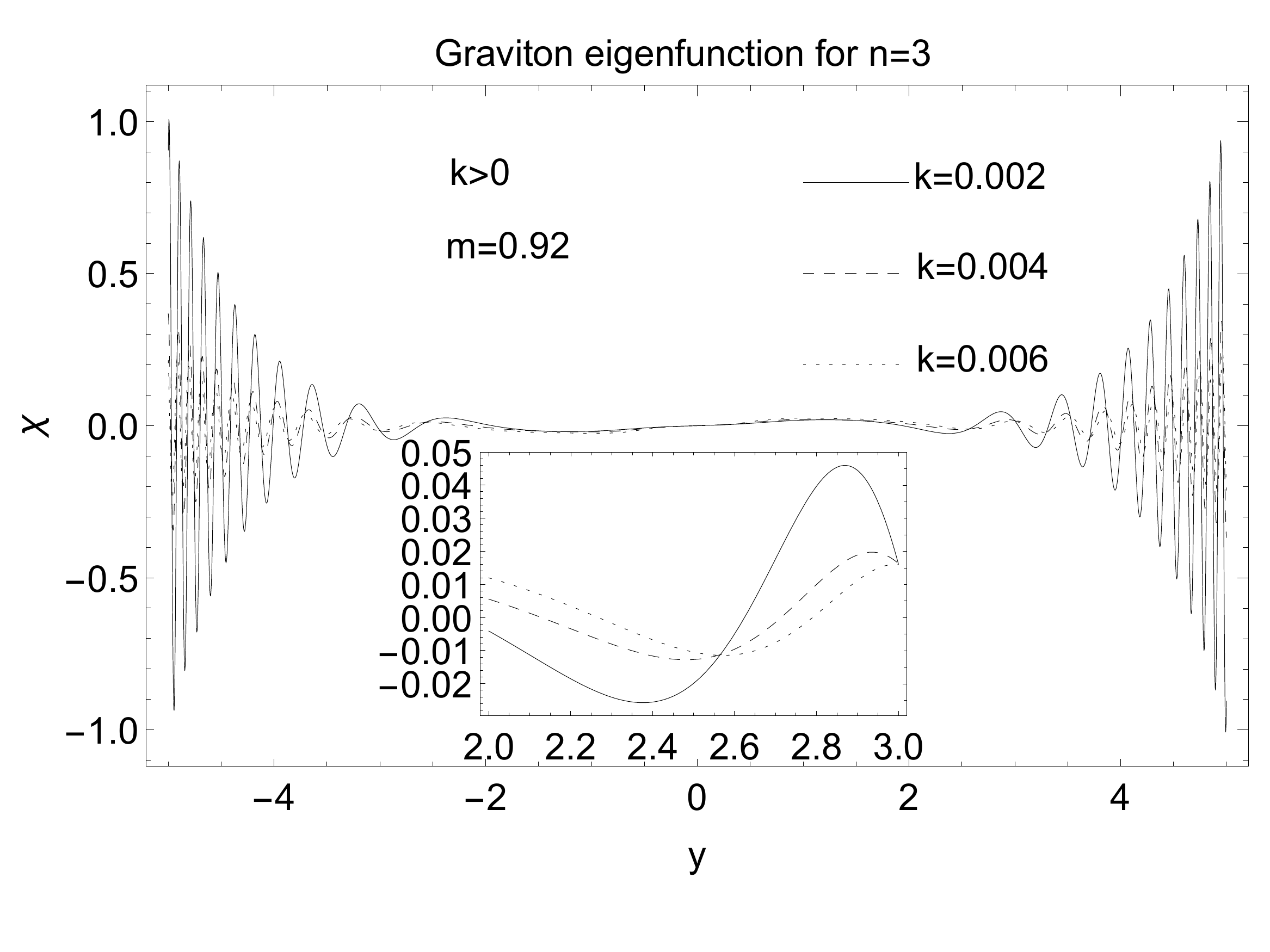}\\
(c) \hspace{8 cm}(d)
\end{tabular}
\end{center}
\caption{Massive modes for $f_1(T,B)$ and $p=\lambda=1$. (a) $n_1=1$.  (b) $n_1=2$ and $k_1=-0.05$. (c) and (d) $n_1=3$ with its first fixed mass eigenfunction.} 
\label{massivemodes}
\end{figure}

\begin{figure}
\begin{center}
\begin{tabular}{ccccccccc}
\includegraphics[height=6cm]{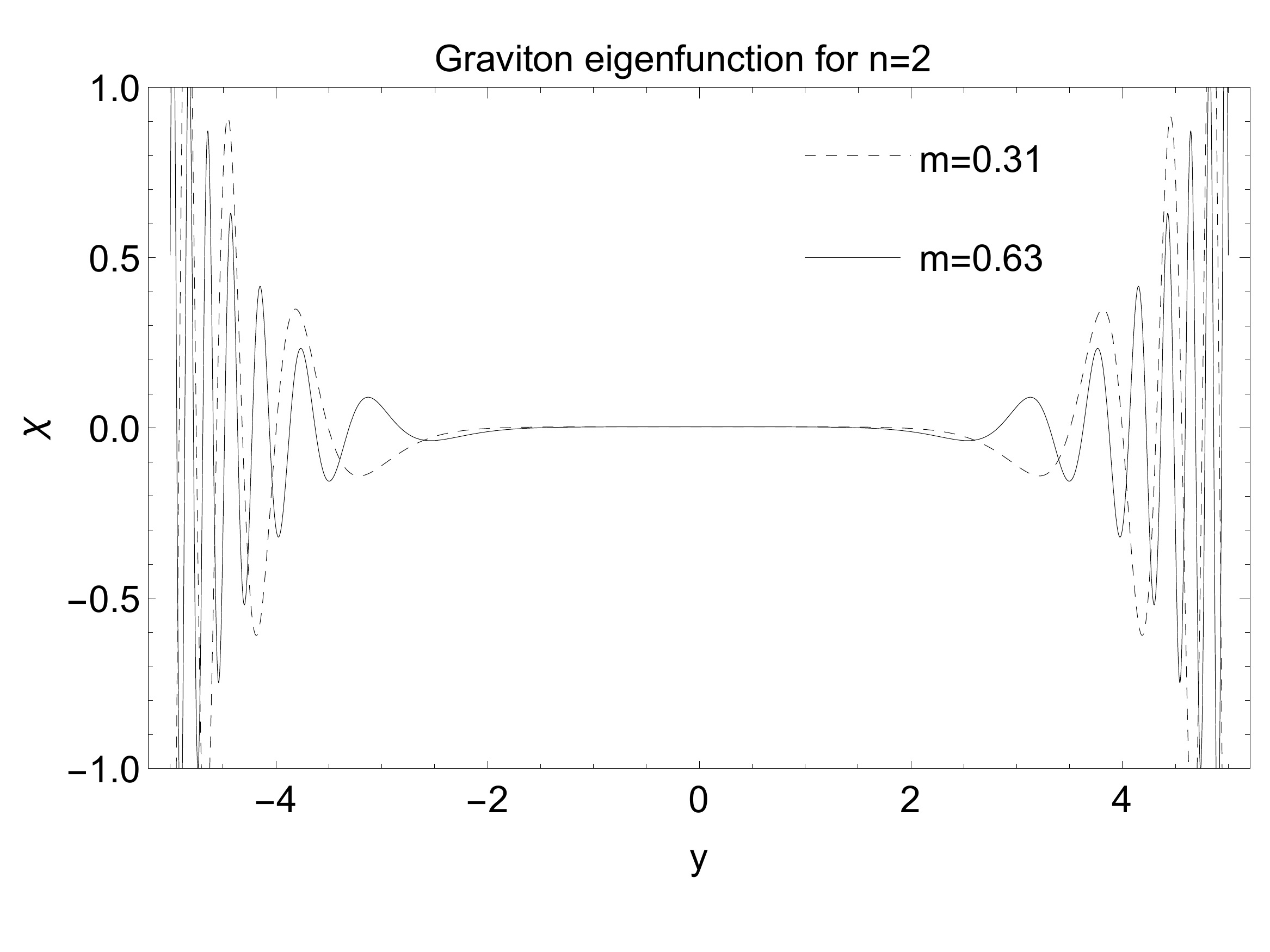} 
\includegraphics[height=6cm]{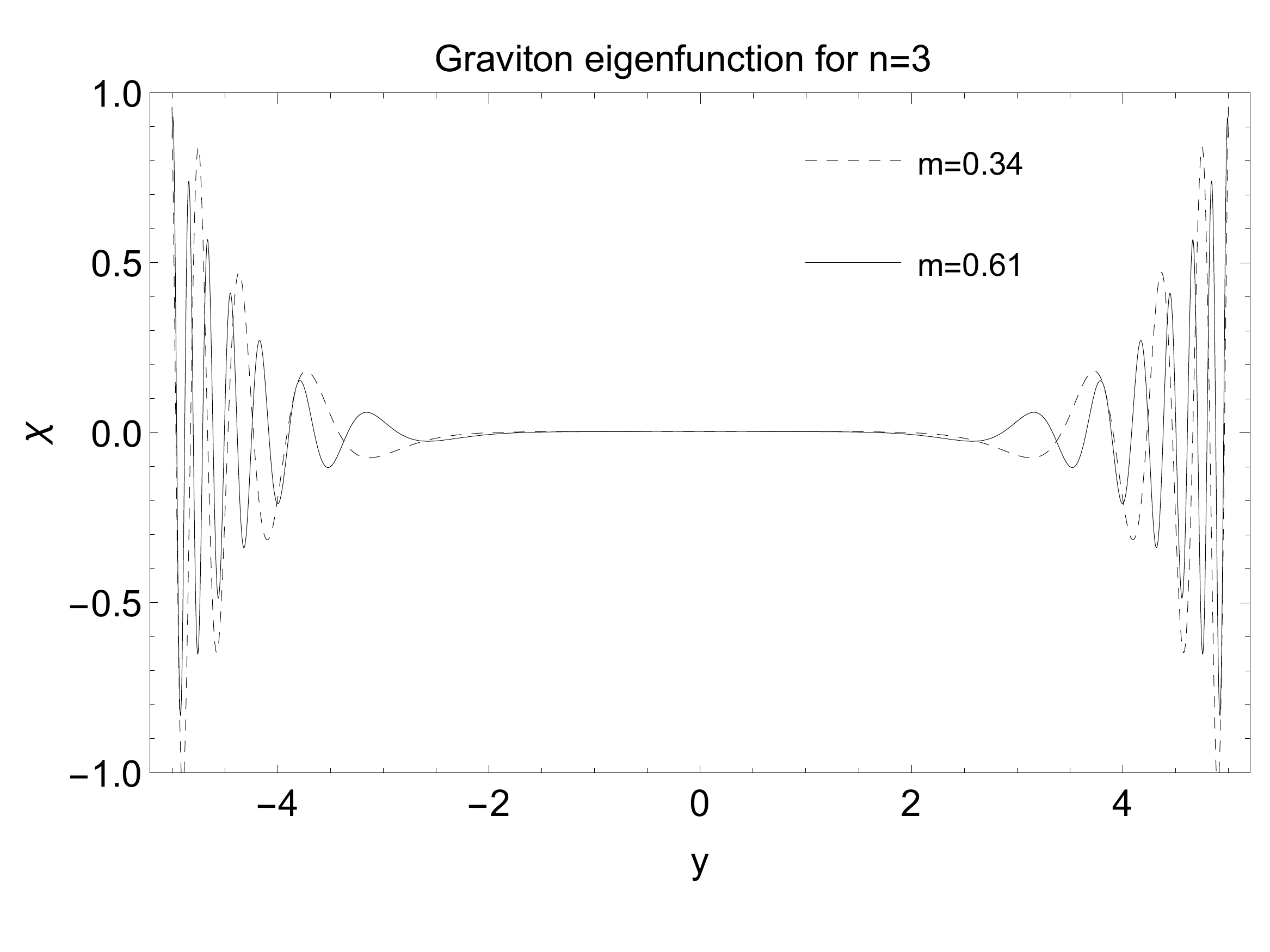}\\
(a) \hspace{8 cm}(b)
\end{tabular}
\end{center}
\caption{Massive modes for $f_2(T,B)$ and $p=\lambda=1$. (a) $n_1=2$. (b) $n_1=3$.} 
\label{massivemodes2}
\end{figure}

On the other hand, for the thick brane regime, we numerically solved the Eq. (\ref{KKequation}) using the interpolation method, thereby obtaining the massive modes for $f_1(T, B)$ and $f_2(T, B)$, adopting the usual boundary condition $\chi'(-\infty)=\chi'(\infty)=0$. As depicted in Fig.\ref{massivemodes} and Fig. \ref{massivemodes2}, the asymptotic divergence of the massive gravitational field shows that they form a tower of non-localized states.

For $f_1(T, B)$ and $n_1=1$, there is no dependency on $k_1$  of the behavior of massive modes as is illustrated in figure \ref{massivemodes} (a). We note that for $n_1=2$ and $3$ is there dependency on $k_1$. It is important to remark that  the amplitude of ripples increase for  $n_1=2$ and an increasing value of $ k_1 $ as well as for $n_1=3$ and a decreasing  value of $k_1$.  The increasing amplitude of the ripples making them more intense and allow their presence within the brane well as illustrated in figures \ref{massivemodes} (c and d), where $n_1=3$ and we varying the parameter $k_1$. For $n_1=2$ the behavior of massive modes is illustrated in figure \ref{massivemodes}(b). 

For $f_2(T, B)$ there is no dependency on $k_2$ in massive modes.  For $n_2=1$ we obtain the same differential equation as $f_1(T, B)$ with $n_1=1$, thus having the same solution and behavior already illustrated in the figure \ref{massivemodes} (a). For any value of $n_2$ there is an amplitude within the brane. For $n_2=2$ the behavior of massive modes is illustrated in figure \ref{massivemodes2}(a) and $n_2=3$ in figure \ref{massivemodes2}(b).

\subsection{Massless modes}
Employing the change to a conformal coordinate $z=\int{e^{-A}}dy$, the tensor perturbation given by Eq.(\ref{KKequation}) is transformed as
\begin{equation}\label{w.17}
\left(\partial_z^2+2H\partial_z+m^2\right)\chi(z)=0,
\end{equation}
where
\begin{eqnarray}\label{34.l}
H=\frac{3}{2}\dot{A}+\frac{4e^{-2A}}{f_T}\Bigg[3\Big(\dot{A}^3-\dot{A}\ddot{A}\Big)(f_{TT}+f_{BT})+\Big(6\dot{A}^3-4\dot{A}\ddot{A}-\dddot{A}\Big)(f_{BB}+f_{TB})\Bigg],
\end{eqnarray}
and the dot denotes differentiation with respect to $z$. With the change on the wave function $\chi(z)=F(z)\Psi(z)$ in Eq.(\ref{w.17}), it is possible the recast into a Sch\"{o}dinger-like equation 
\begin{eqnarray}\label{36.l}
[-\partial_z^2+U(z)]\Psi(z)=m^2\Psi(z),
\end{eqnarray}
where the potential is defined by
\begin{eqnarray}\label{potential}
U(z)=\dot{H}+H^2,
\end{eqnarray}
and 
\begin{eqnarray}
F(z)=e^{-\frac{3}{2}A+\int K(z)dz},
\end{eqnarray}
with
\begin{eqnarray}
\label{kz}
 K(z)=-\frac{4e^{-2A}}{f_T}\Bigg[3\Big(\dot{A}^3-\dot{A}\ddot{A}\Big)(f_{TT}+f_{BT})+\Big(6\dot{A}^3-4\dot{A}\ddot{A}-\dddot{A}\Big)(f_{BB}+f_{TB})\Bigg].
\end{eqnarray}
The Schrödinger-like Eq.(\ref{36.l}) can be factorized as
\begin{eqnarray}
\left(-\partial_z+H\right)\left(\partial_z+H\right)\Psi(z)=m^2\Psi(z),
\end{eqnarray}
which represents an equation of the so-called supersymmetric quantum mechanic. The superpotential $H$ and the quantum mechanic supersymmetric form of the potential $U$ ensures the absence of tachyonic KK gravitational modes. 

Besides the spectrum stability, the potential in Eq. (\ref{potential}) allows a massless KK mode of the form 
\begin{eqnarray}
\Psi_0=N_0e^{\frac{3}{2}A-\int K(z)dz},
\end{eqnarray}
where $N_0$ is a normalization constant. In order to recover the four-dimensional gravity, the zero mode should be localized on the brane. Let us now investigate the problem of localization of the massless mode.

\begin{figure}
\begin{center}
\begin{tabular}{ccccccccc}
\includegraphics[height=5cm]{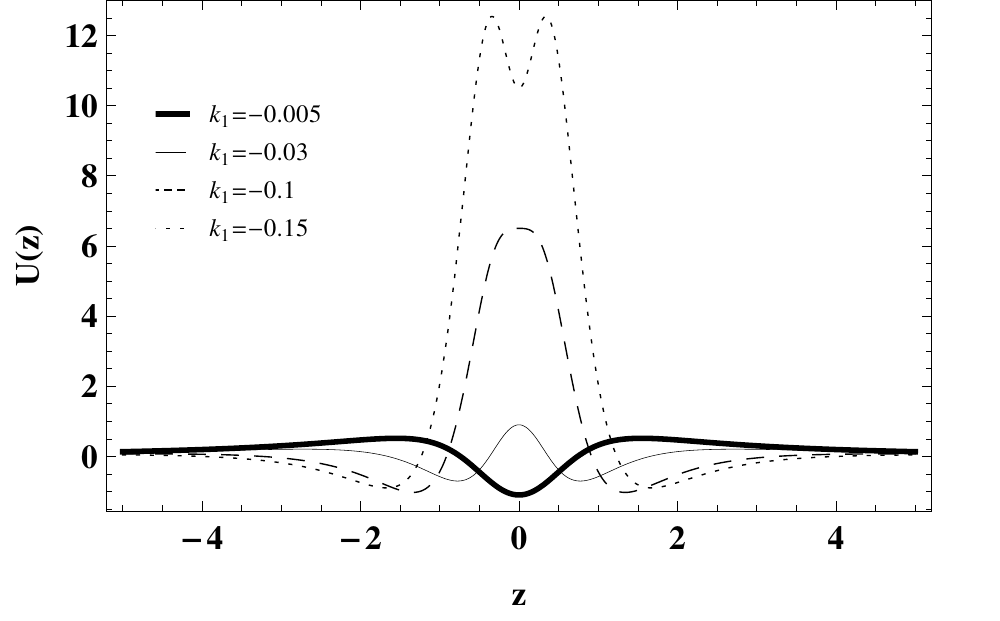}
\includegraphics[height=5cm]{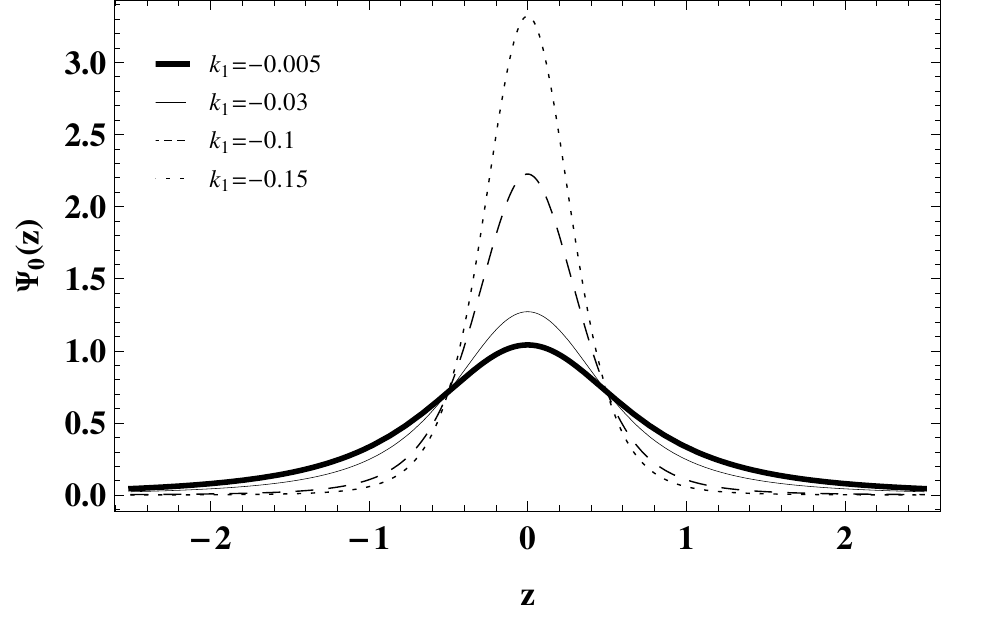}\\
(a) \hspace{8 cm}(b)\\
\includegraphics[height=5cm]{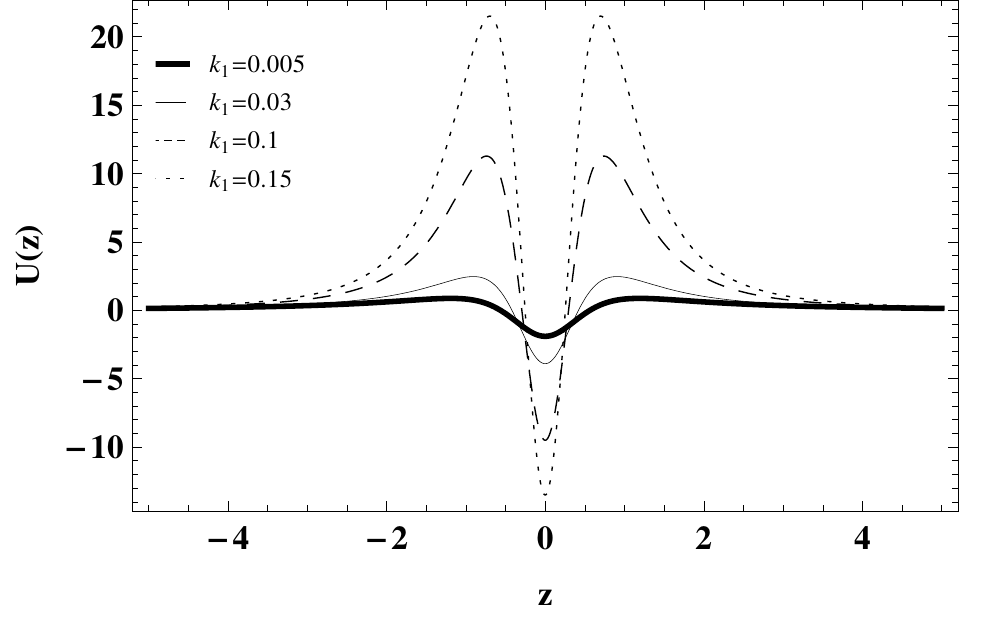} 
\includegraphics[height=5cm]{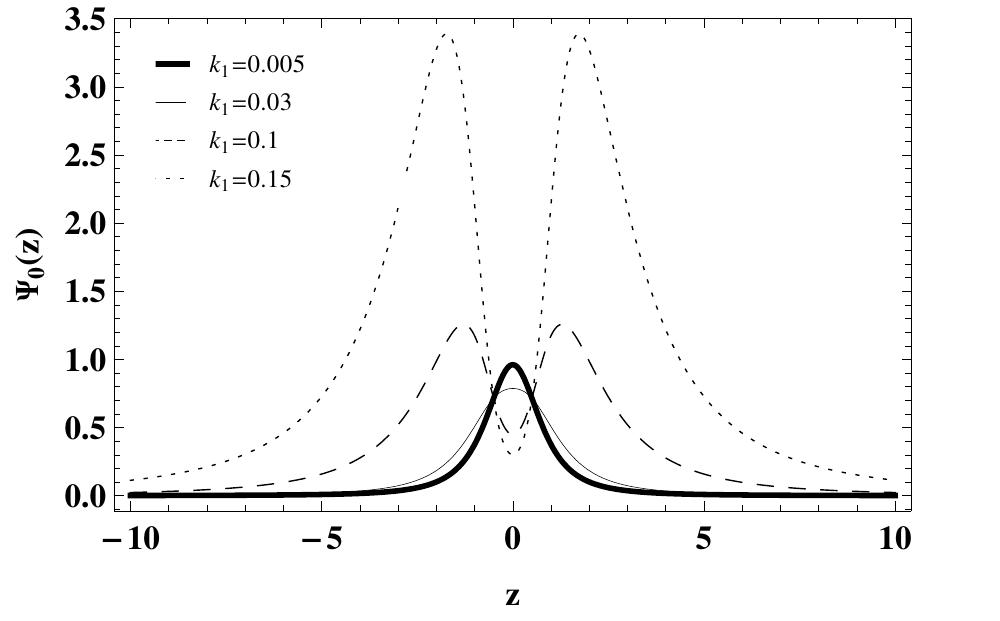}\\
(c) \hspace{8 cm}(d)
\end{tabular}
\end{center}
\caption{Plots of the effective potential, and zero mode for $n_1=2$ and $p=\lambda=1$. (a) and (b) $k_1<0$. (c) and (d) $k_1>0$.
\label{figPE1}}
\end{figure}

\begin{figure}
\begin{center}
\begin{tabular}{ccccccccc}
\includegraphics[height=5cm]{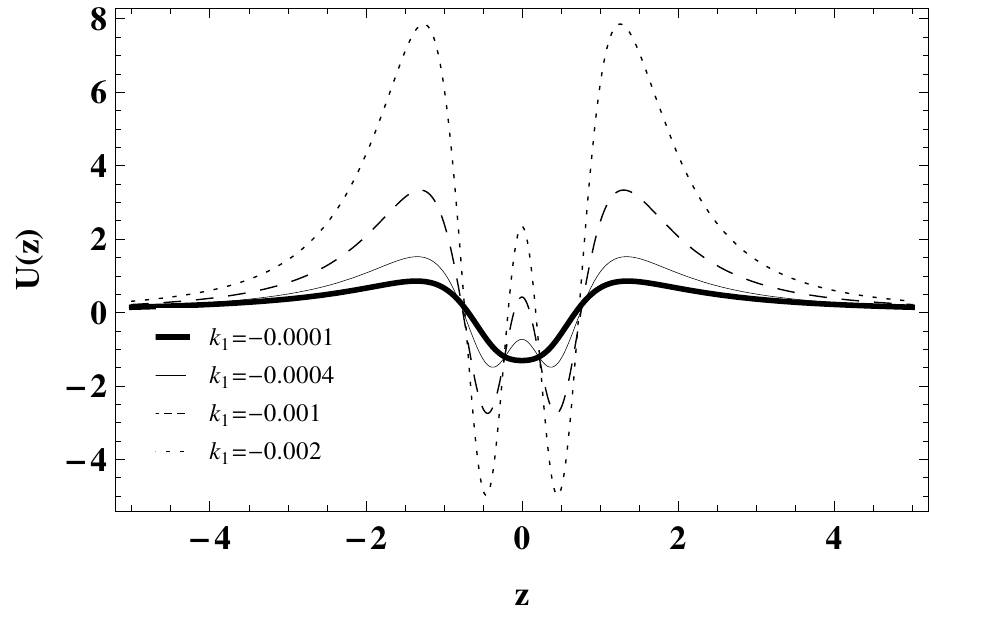}
\includegraphics[height=5cm]{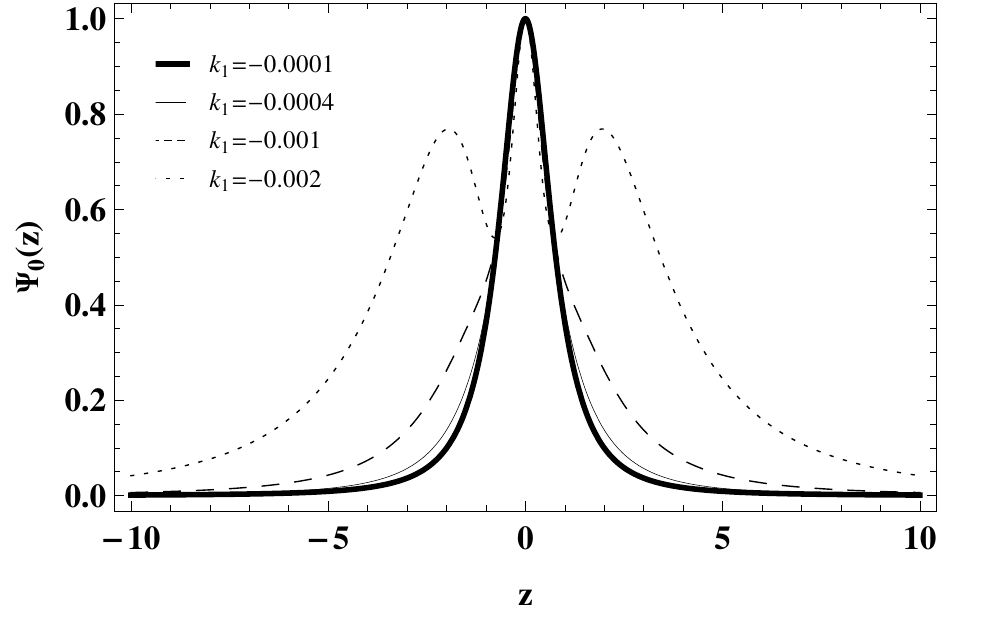}\\
(a) \hspace{8 cm}(b)\\
\includegraphics[height=5cm]{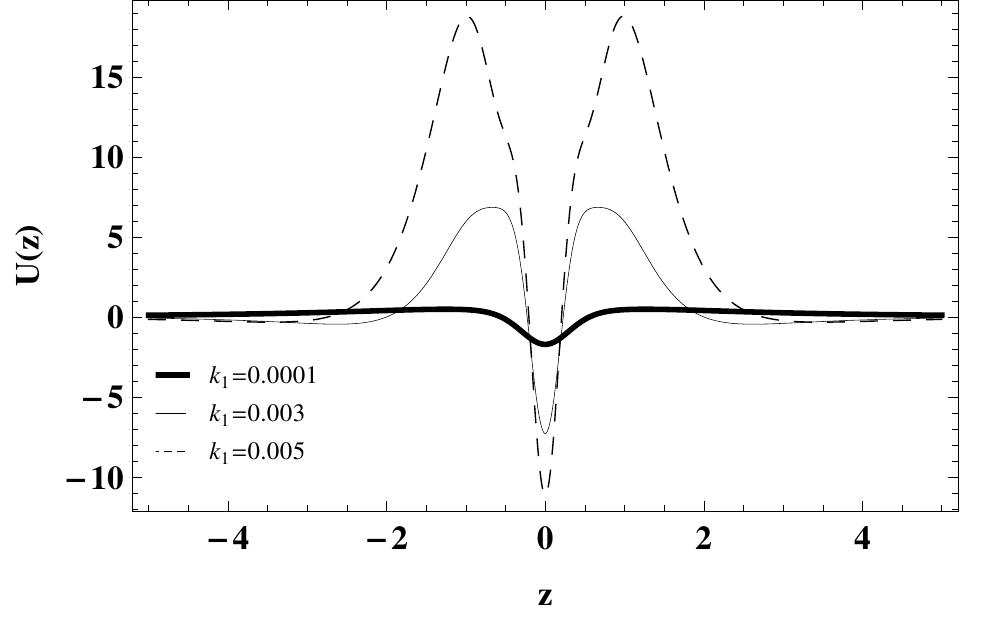} 
\includegraphics[height=5cm]{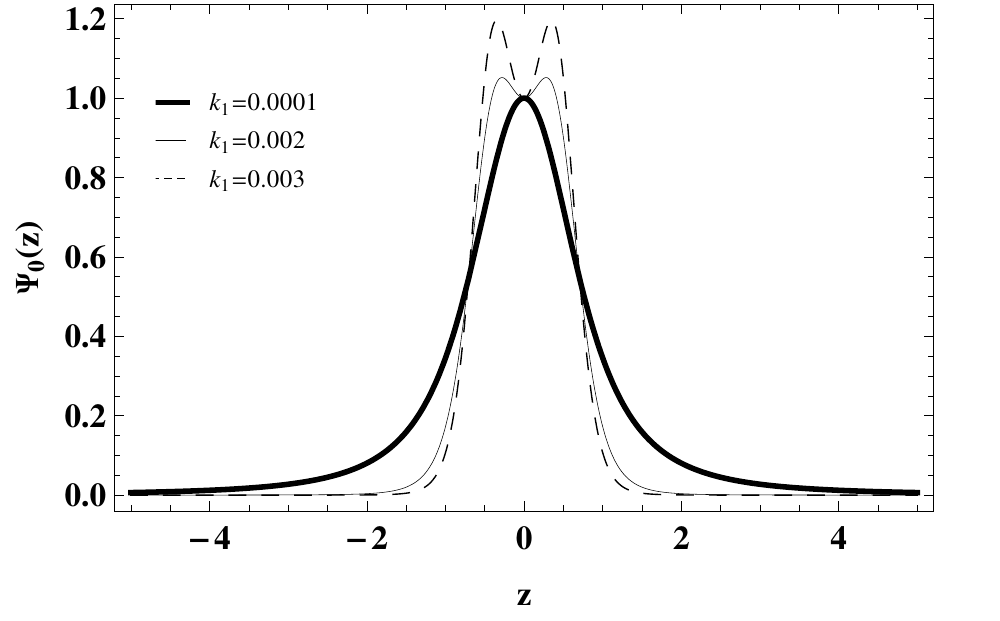}\\
(c) \hspace{8 cm}(d)
\end{tabular}
\end{center}
\caption{Plots of the effective potential, and zero mode for $n_1=3$ and $p=\lambda=1$. (a) and (b) $k_1<0$. (c) and (d) $k_1>0$. 
\label{figPE2}}
\end{figure}

\begin{figure}
\begin{center}
\begin{tabular}{ccccccccc}
\includegraphics[height=5cm]{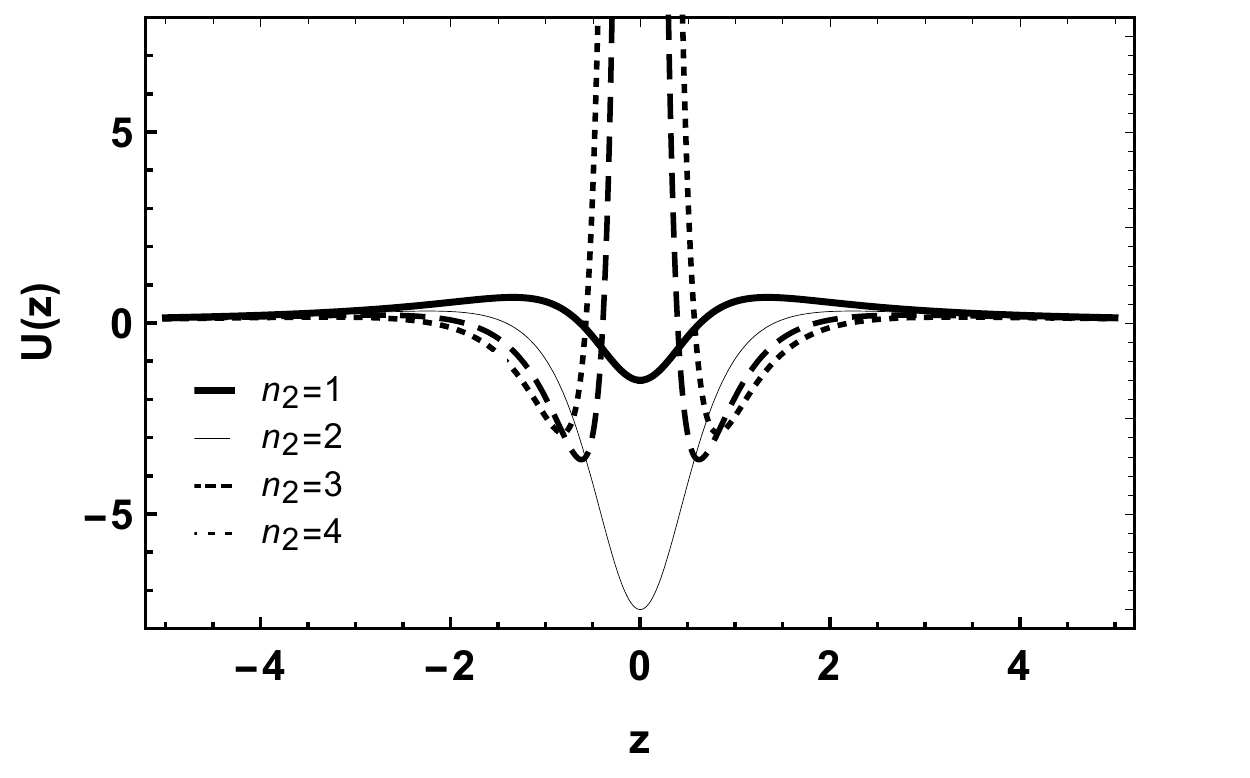}
\includegraphics[height=5cm]{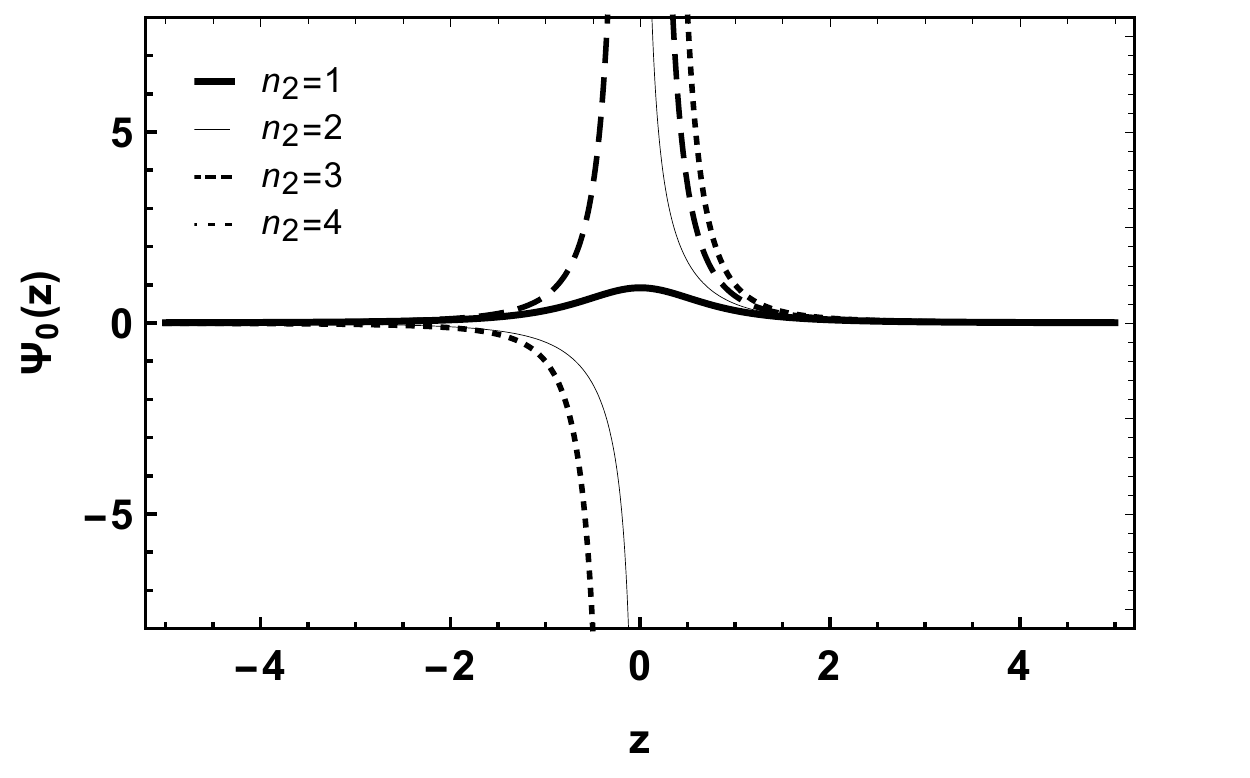}\\
(a) \hspace{8 cm}(b)
\end{tabular}
\end{center}
\caption{Plots of the effective potential (a). Zero mode (b). For  $p=\lambda=1$. 
\label{figPE3}}
\end{figure}

For $f_1(T,B)$  the superpotential (\ref{34.l}) has the form
\begin{eqnarray}\label{43.l}
H_1&=&-\frac{3pz\lambda^2}{\xi}-8^{n_1-1}k_1n_1(n_1-1)zp\lambda^4\xi^{p-3}[(p\lambda)^2(3z^2-1)+2p\xi+3]\nonumber\\
&\times&\Big\{ p\lambda^2\xi^{p-2}[1-(1-4p)(z\lambda)^2] \Big\}^{n_1-2},
\end{eqnarray}
where $\xi\equiv 1+(z\lambda)^2$. The expression of the potential is too lengthy to be written here. Instead, we plot the potential and zero mode and explore some qualitative features. Figures \ref{figPE1} and \ref{figPE2} show us how the potential and the zero mode recognize the division of the brane by varying the parameters that control the boundary term of $f_1(T,B)$.

For $n_1=1$ the effective potential is independent of the parameter $k_1$. For $n_1=2$ and $k<0$ as we can see from Fig.\ref{figPE1}(a), when $k_1$ decreases, the shape of the effective potential is volcano like, and the zero mode wave function has only one peak getting more localized as seen in Fig.\ref{figPE1}(b).  When $k_1>0$ increases, as we can see from Fig.\ref{figPE1}(c), two new potential barrier appear away from the origin and the potential well around the origin will increases. As a result, the zero mode wave function splits into two peaks as shown in Fig.\ref{figPE1}(d).

For $n_1=3$ and $k<0$ as we can see from Fig.\ref{figPE2}(a), when $k_1$ is decreasing, the potential well splits in two around the origin and two new potential barrier, and the zero mode wave function divides into three, with a peak at the origin and the other two further away from the origin as seen in Fig.\ref{figPE2}(b).  When $k_1>0$ increases, as we can see from Fig.\ref{figPE2}(c), two new potential barrier appear away from the origin and the potential well around of the origin will increases. As a result, the zero mode wave function splits into two peaks as shown in Fig.\ref{figPE2}(d).
 
For $f_2(T,B)$  the superpotential (\ref{34.l}) has the form
\begin{eqnarray}\label{43.l}
H_2=-\frac{(n_2-1)}{\xi}+\frac{z\lambda^2}{2\xi}\Big[2+2n_2(p-1)-5p\Big].
\end{eqnarray}
 
As we can see from Eq.(\ref{43.l}), there is no dependency on the $k_2$ parameter. The figure \ref{figPE3} show us how the potential and the zero mode recognizes the division of the brane by varying the parameters that control torsion term of $f_2(T,B)$. When $n_2=1$ we have the same solution as $f_1(T,B)$ with $n_1=1$. As we can see in the Fig.\ref{figPE3}(a), the potential goes from a well to a delta-type barrier when we increase the parameter $n_2$. As a result, zero modes become non-localized for values of $n_2=2,4,...$ (even numbers) as shown in Fig.\ref{figPE3}(b).

\section{Final remarks}
\label{sec3}

We studied the torsion and  boundary term effects on a braneworld in the context of the $f(T,B)$ modified teleparallel gravity. For this, we propose two particular cases for $ f(T,B)$, namely $f_1(T,B)=T+k_1B^{n_1}$ and $f_2(T,B)=B+k_2T^{n_2}$.We note that linear term for $B$ in $f_2$ may be omitted as it does not contribute to the equations of motion, in this case $f_2$ is equivalent to a $f(T)$ case where $f(T)=kT^{n}$. In both cases the torsion and  boundary term produces an inner brane structure tending to split the brane. Furthermore, the $f(T,B)$ modified the exterior region making the solutions depend on the parameters that control the torsion and  boundary term $n_ {1,2}$ and $k_ {1,2}$. Even with the cosmological constant being zero, for $ f_1 (T, B) $ it was possible to obtain solutions. 

The vaccum expectation value and the profile of the scalar field inside the core are controlled by the parameters that control torsion and boundary term. The profile of the scalar field suggests a topological stability for $f_1(T,B)$. However for $f_2(T,B)$ the stability is only observed for the values of the parameter  $n_2=1, 3, 5,...$ odd. The thick brane undergoes a phase transition evinced by the energy density components. Similar behavior was found for $f(T)$ in Refs \cite{Yang2012, tensorperturbations}. As the  parameters $n_ {1,2}$ and $k_ {1,2}$ increase, the source violates the dominant energy condition, which reflects on the negative density responsible for the brane splitting. 

For $f_1(T,B)$  with the first massive values, the massive mode is dependent on the parameters that control the boundary term, well evidenced for $ n_1 = 3 $ where  decreasing the value of $k_1$, increases the amplitude of the ripples making them more intense and 
presenting ripples within the brane. For $ f_2 (T, B) $ with the first massive values, the massive mode is dependent on the parameter $ n_2 $ that control the torsion term and presents no amplitude within the brane. Therefore, the brane splitting process leads to modifications of the massive gravitons inside the thick brane.  
The interaction of the massive modes with the torsion and boundary term is more intense inside the brane core where the amplitude and the rate of growth depend on the parameters $n_{1,2}$ and $k_{1,2}$.

The analysis of the Schr\"{o}dinger-like potential reveals the effects of the torsion and boundary term on the KK modes. For $f_1(T,B)$    with $n_1=2$, increasing $k_1$, two new potential barrier appear away from the origin and the potential well around of the origin will increases. As a result, the zero mode wave function splits into two peaks. For $n_1=3$, when $k_1$ decreasing, the potential well splits in two around the origin and two new potential barrier, and the zero mode wave function divides into three, with a peak at the origin and the other two further away from the origin.  When $k_1$ increases, two new potential barrier appear away from the origin and the potential well around of the origin will increases. As a result, the zero mode wave function splits into two peaks. We find an interesting configuration for $f_2(T,B)$ where the potential goes from a well to a delta-type barrier when we increase the parameter $n_2$. As a result, zero modes become non-localized for values of $n_2=2,4,...$ even.

\section*{Acknowledgments}
\hspace{0.5cm}The authors thank the Conselho Nacional de Desenvolvimento Cient\'{\i}fico e Tecnol\'{o}gico (CNPq), grants n$\textsuperscript{\underline{\scriptsize o}}$ 312356/2017-0 (JEGS) and n$\textsuperscript{\underline{\scriptsize o}}$ 308638/2015-8 (CASA), and Coordena\c{c}\~{a}o de Aperfei\c{c}oamento do Pessoal de N\'{i}vel Superior (CAPES), for financial support. The authors also thank the anonymous referee for their valuable comments and suggestions.


\end{document}